\documentclass[11pt,preprint,twocolumn]{emulateapj}
\singlespace
\newcommand{\kms}{\,{\rm km\,s^{-1}}}
\newcommand{\kmss}{\,{\rm km^2\,s^{-2}}}
\newcommand{\msun}{\,{\rm M_\odot}}
\newcommand{\msuny}{\,{\rm M_\odot}{\rm yr}^{-1}}

\newcommand{\vkick}{{V_{\rm kick}}}

\def\spose#1{\hbox to 0pt{#1\hss}}
\def\lta{\mathrel{\spose{\lower 3pt\hbox{$\mathchar"218$}} \raise 2.0pt\hbox{$\mathchar"13C$}}}
\def\gta{\mathrel{\spose{\lower 3pt\hbox{$\mathchar"218$}} \raise 2.0pt\hbox{$\mathchar"13E$}}}

\usepackage{multirow}

\begin{document}
\title{Recoiling massive black holes in gas-rich galaxy mergers}
\author{Javiera Guedes\altaffilmark{1}, Piero Madau\altaffilmark{1}, Lucio Mayer\altaffilmark{2}, \& Simone Callegari\altaffilmark{2}}

\altaffiltext{1}{Department of Astronomy \& Astrophysics, University of California, Santa Cruz, CA 95064.}
\altaffiltext{2}{Institute for Theoretical Physics, University of Zurich, Winterthurerstrasse 190, CH-9057 Zurich, Switzerland.}

\begin{abstract}

The asymmetric emission of gravitational waves produced during the coalescence of a massive black hole (MBH) binary imparts a velocity ``kick" to the 
system that can displace the hole from the center of its host. Here we study the trajectories and observability of MBHs recoiling in three (one major, two minor) 
gas-rich galaxy merger remnants that were previously simulated at high resolution, and in which the pairing of the MBHs had been shown to be successful. We run new 
simulations of MBHs recoiling in the major merger remnant with Mach numbers in the range $1\le \mathcal{M} \le 6$, and use simulation data to construct a semi-analytical 
model for the orbital evolution of MBHs in gas-rich systems. We show that: 1) in major merger remnants the energy deposited by the moving hole into the rotationally 
supported, turbulent medium makes a negligible contribution to the thermodynamics of the gas. This contribution becomes significant in minor merger remnants, 
potentially allowing for an electromagnetic signature of MBH recoil; 2) in major merger remnants, the combination of both deeper central potential well and 
drag from high-density gas confines even MBHs with kick velocities as high as $1,200\,\kms$ within 1 kpc from the host's center; 3) kinematically offset nuclei may be 
observable for timescales of a few Myr in major merger remnants in the case of recoil 
velocities in the range $700-1,000\,\kms$; 4) in minor mergers remnants the effect of gas drag is weaker, and MBHs with recoil speeds in the range $300-600\,\kms$ will 
wander through the host halo for longer timescales. When accounting for the probability distribution of kick velocities, however, we find that the likelihood of observing 
recoiling MBHs in gas-rich galaxy mergers is very low, typically below $10^{-5}-10^{-6}$.
\end{abstract}


\section{Introduction}

The pairing of massive black holes  (MBHs) is a natural consequence of galaxy mergers in hierarchical structure formation scenarios \citep{begelman80,volonteri03}. To date, over 
30 dual active galactic nuclei (AGNs) have been discovered in merger remnants at redshifts $0.34<z<0.82$ in the DEEP Survey \citep{comerford09a}, over 160 have be 
reported from studies of double-peaked [OIII] profiles \citep[e.g.][]{zhou04,gerke07,liu10,smith10} and nearly 40 from the Swift BAT survey \citep{koss10}. Furthermore,
\textit{Chandra} observations of the nucleus of NGC6240 have unveiled a MBH binary with separation $<1$ kpc \citep{komossa03, max07}. The coalescence of MBH binaries 
may become the dominant source of gravitational waves at mHz frequencies \citep[e.g.,][]{haehnelt94,wyithe03,Sesana04,Sesana05}, a sensitivity range probed by the 
\textit{Laser Interferometer Space Antenna} (LISA).  

Asymmetries in the configuration of the coalescing binary cause the beaming of gravitational wave radiation in a preferred direction. To conserve linear momentum, the remnant 
hole recoils in the direction opposite to the radiative flux \citep{Peres62,Bekenstein73,Fitchett84}. The recoil or ``kick" velocity, $\vkick$, depends on the binary 
mass ratio and on the magnitude and direction of their spins, and does not depend on the total mass of the binary. Recent advances in numerical relativity 
\citep{Pretorius05,Campanelli06} have allowed several groups \citep{Baker06,Herrmann07,Gonzalez2007} to evolve binary black holes from inspiral through coalescence
and estimate the resulting recoil velocities as a function of mass ratio and spin \citep{baker08,Campanelli2007}. These studies show kick velocities in the range from a 
hundred to a few thousand $\kms$. The maximum speed, $V_{\rm max} = 3750~\kms$, is obtained for equal-mass maximally-spinning black holes with anti-aligned spin vectors
in the orbital plane of the binary \citep{Campanelli07}. 

When $\vkick$ is lower than the escape speed of the host, the MBH will wander through the galaxy in an orbit that depends on the detailed characteristics of 
the host potential. Several groups have calculated the trajectories of MBHs in smooth spherical potentials \citep{madau04,gualandris08,blecha08}. Asphericities in the 
dark matter potential result in highly non-radial MBH orbits, increasing its decay timescale compared to the spherical case 
\citep{vicari07,guedes08b,guedes09}. In gas-rich galaxy mergers, the wandering hole may travel through a rotationally supported gaseous disk that is inhomogeneous, 
non-axisymmetric, clumpy, and turbulent. These factors affect the kinematics of the recoiling MBH and its ability to accrete new fuel. In this paper, we 
study the orbits of recoiling MBHs in high-resolution galaxy merger remnants that have successfully formed black hole pairs. These remnants are the endpoint of 
the 1:1 major merger simulation of \cite{mayer07} and the 1:4 and 1:10 minor merger simulations of \cite{callegari09}. We construct a semi-analytical model based on the 
host potential formed after the merger and calculate the trajectories of MBHs for arbitrary recoil velocities and inclination angles, accounting for the effect of dynamical 
friction and gas drag. This allows us to obtain the characteristic apocenter distances and return times for MBHs recoiling in different environments. In addition, in order
to study the possible effect of the moving hole on the ambient gas, we perform high-resolution $N$-body + SPH simulations of recoiling MBHs in the major merger galaxy remnant.
Under the assumption that the MBH binary is able to coalesce efficiently after reaching the minimum length scale resolved by the simulation, the hole is given a kick and
launched into orbit in an environment characterized by a turbulent, actively accreting gaseous disk surrounded by unrelaxed stellar and dark matter components. 

The paper is organized as follows. The probability distribution of kick velocities is reviewed in \S~\ref{kickvelocity}. We describe the main properties of our 
three host galaxies in \S~\ref{hosts}, and our semi-analytical model in \S~\ref{model}. We assess the response of the gaseous disk to the motion of the MBH, 
together with its detectability as a velocity-offset or spatially-offset AGN in \S~\ref{detectability}. Finally, we discuss and summarize our results in \S ~\ref{summary}.

\section{Distributions of kick velocities}\label{kickvelocity}

\begin{figure}[th]
\centering
\includegraphics[width=.5\textwidth]{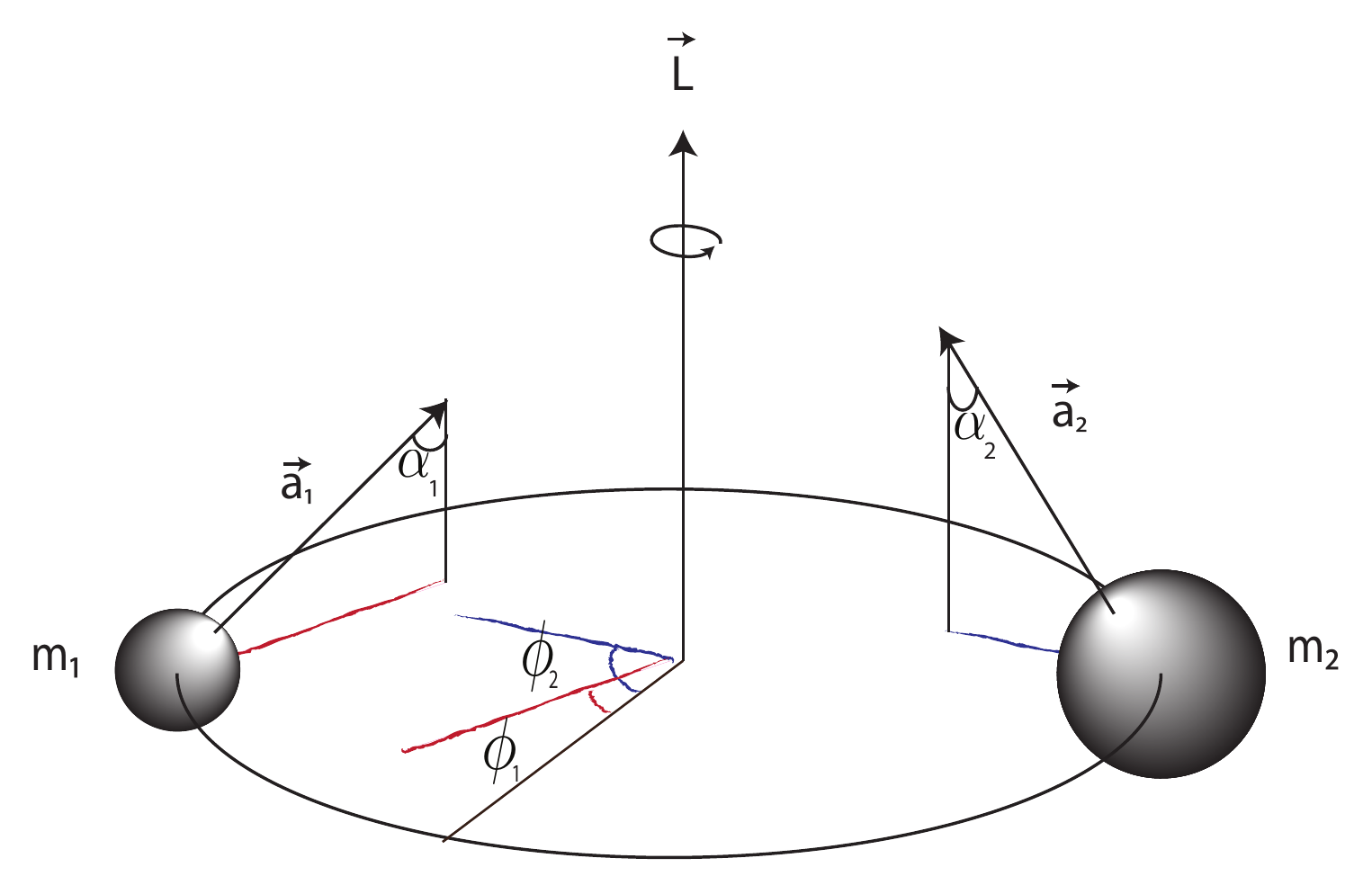}
\caption{Geometry of the massive black hole binary prior to coalescence.}  
\label{schematic}
\end{figure}

The recoil velocity depends on the binary mass ratio, $q=m_1/m_2 \le 1$, on the magnitude of the black hole spin vectors $\vec{a}_{1,2} = \vec{S}_{1,2}/m_{1,2}$, and on 
their orientation with respect to the orbital angular momentum vector. Fitting formulae for $\vkick$ to fully general relativistic numerical calculations, given a 
particular binary configuration, have been provided by several authors \citep[e.g.][]{Campanelli07,baker08,vanMeter10}. In the formulation of \cite{vanMeter10}:
\begin{eqnarray}
\vec{V}_{\rm kick} &=& v_{\perp,m} \, \vec{e}_x + v_{\perp,s} (\cos\xi \, \vec{e}_x + \sin\xi \, \vec{e}_y)
+ v_{\parallel} \, \vec{e}_z, \\
v_{\perp,m} &=& A \eta^2 \sqrt{1-4\eta}\, (1 + B \eta), \\
v_{\perp,s} &=& H \frac{\eta^2}{(1+q)}\, ( a_2 \cos{\alpha_2} - q\, a_1 \cos{\alpha_1}), \\
v_{\parallel} &=& \frac{K_2 \eta^2 + K_3 \eta^3}{1+q}  [ q\, a_1 \sin{\alpha_1}\cos(\phi_1 - \Phi_1) \label{vpara} \\ 
\nonumber			 & & - a_2 \sin{\alpha_2}\cos(\phi_2 -\Phi_2) ] \\
\nonumber			 &+& \frac{K_S (q-1) \eta^2}{(1+q)^3}  [ q^2\, a_1 \sin{\alpha_1}\cos(\phi_1 - \Phi_1) \\	
\nonumber			 & & + a_2 \sin{\alpha_2}\cos(\phi_2 -\Phi_2) ],
\end{eqnarray}
where $v_{\perp,m}$ is the contribution to the recoil velocity from mass asymmetry, the indices ${\parallel}$
and ${\perp}$ refer to projections parallel and perpendicular to the orbital angular momentum, and
$\vec{e}_x$ and $\vec{e}_y$ are orthogonal unit vectors in the orbital plane. Here $a_{1,2}$ are the magnitudes of the hole spin vectors, $\alpha_{1,2}$ the 
angles between $\vec{a}_{1,2}$ and the orbital angular momentum vector $\vec{L}$, $\eta \equiv q/(1+q)^2$ is the symmetric mass ratio, $\phi_{1,2}$ are 
angles between the projection of $\vec{a}_{1,2}$ along the orbital plane and a reference angle, and $\Phi_{1,2}$ are constants that depend on the mass ratio. 
Figure~\ref{schematic} shows a schematic of the binary geometry prior to coalescence. The best-fit parameters are $A = 1.35 \times 10^4 \kms$, $B = -1.48$, 
$H = 7,540 \kms$, $\xi =215^\circ$, $K_2 = 3.21\times 10^4 \kms$, $K_3 = 1.09\times 10^5 \kms$, and $K_S = 1.54\times 10^4 \kms$ \cite{vanMeter10}.
%
\begin{figure}
\centering
\includegraphics[width=.4\textwidth]{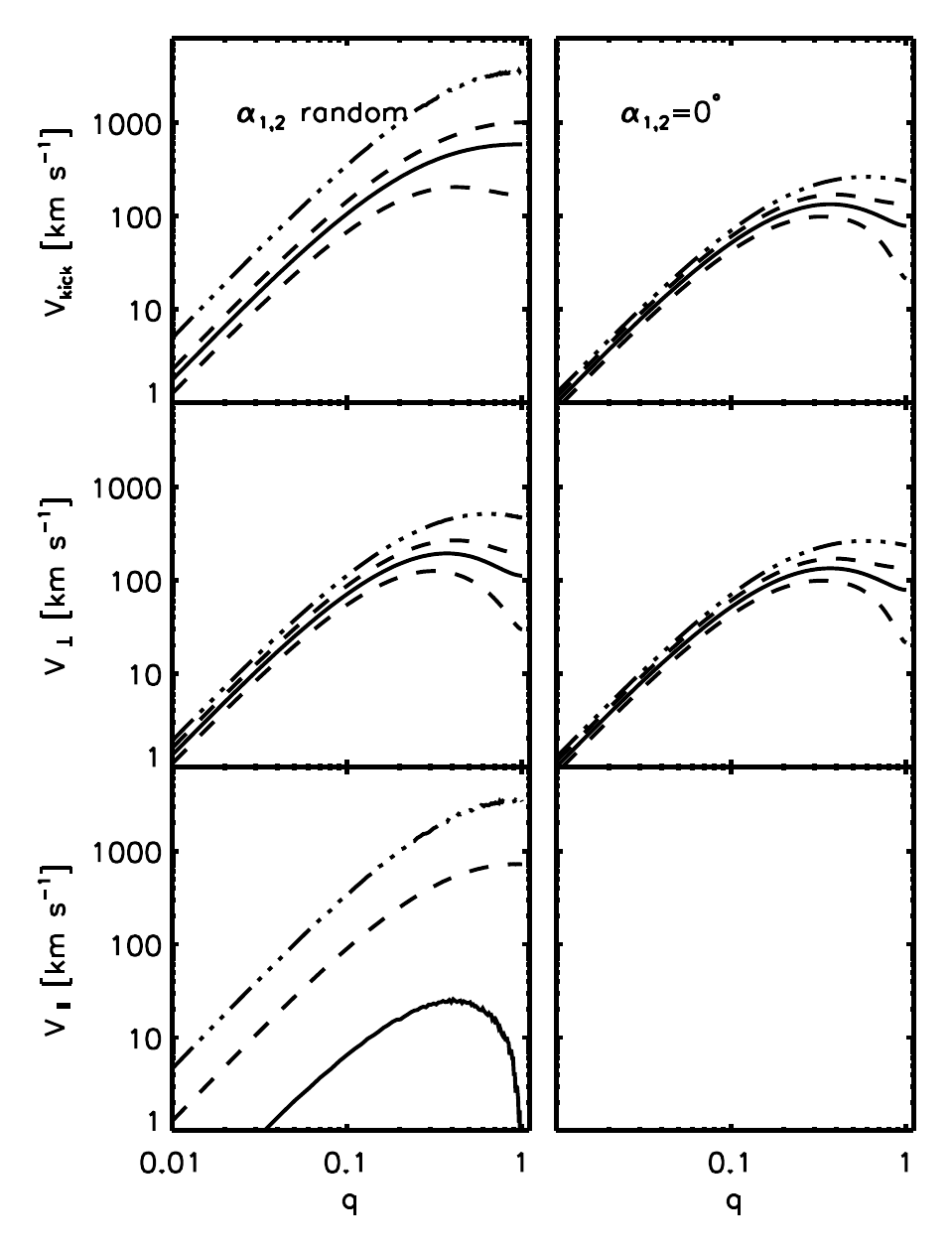}
\caption{Distribution of recoil velocities as a function of binary mass ratio. {\it Left panel:}  randomly distributed spin orientations. {\it Right panel:} 
aligned spins oriented along the orbital angular momentum vector. In both cases the spin magnitudes are randomly sampled from an uniform distribution 
in the range $0 \le a_{1,2}\le 1$. The solid line represents the mean of the distribution, the dashed lines show 1$\sigma$ deviations from the mean, and 
the dash-dotted line shows the maximum value attainable. The component $v_{\parallel}$ vanishes in the case $\alpha_{1,2}=0$.}  
\label{distributions}
\end{figure}

To obtain the recoil velocity distribution as a function of $q$ and the probability distribution function at a given $q$, we sample the equation above using a 
method similar to that of \cite{tanaka09}. For each value of the binary mass ratio we carry out $10^6$ realizations where the black hole spins have random 
orientations (with $0\le \alpha_{1,2}\le \pi$)
or are aligned with the angular momentum vector ($\alpha_{1,2}=0$). The angle $\phi_{1,2} -\Phi_{1,2}$ and spin magnitudes are randomly sampled from a uniform 
distribution in the range $0-2\pi$ and $0-1$, respectively.
%
%
\begin{figure}
\centering
\includegraphics[width=.4\textwidth]{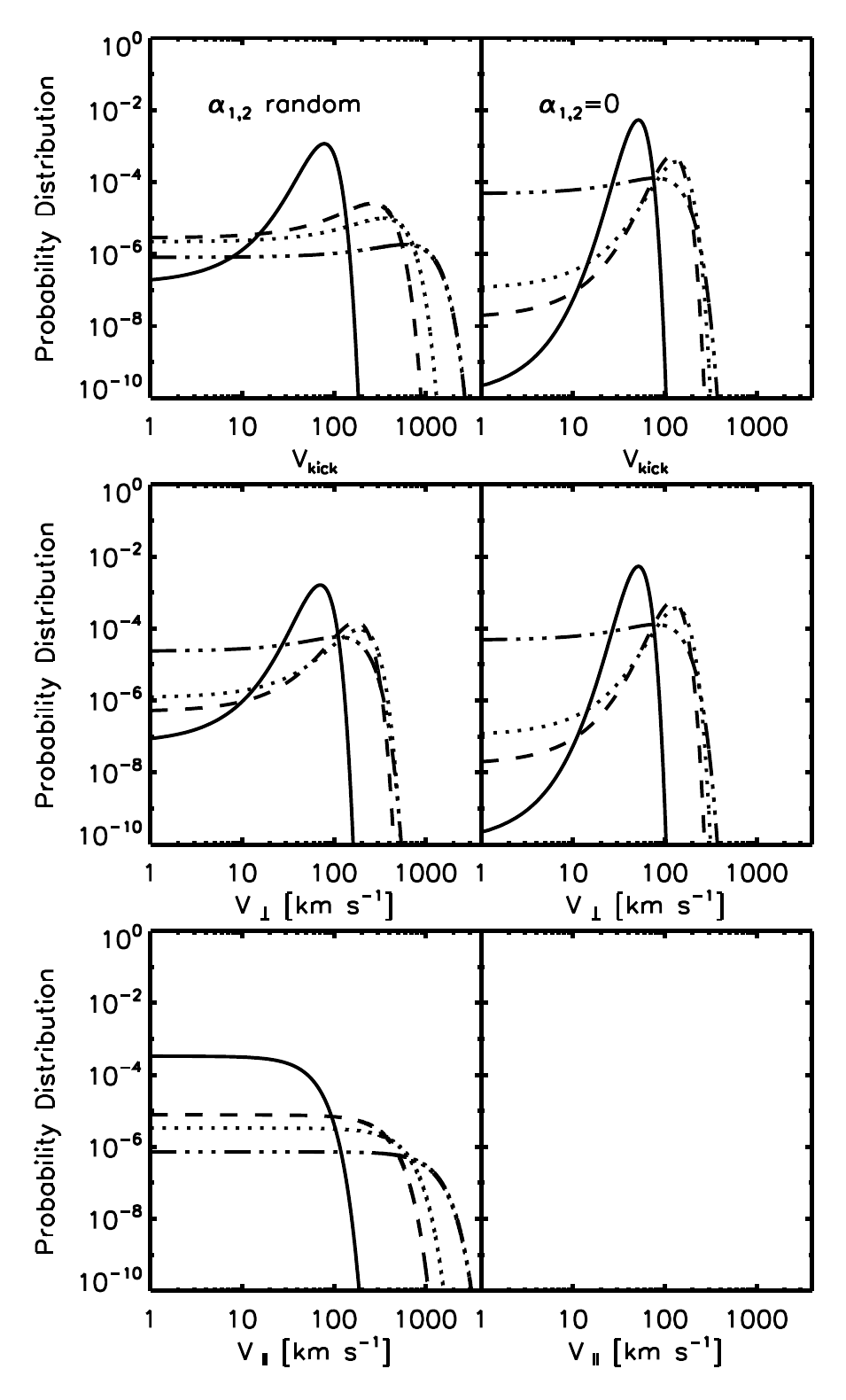}
\caption{Probability distribution function (PDF) of total recoil velocities and their components parallel and perpendicular to the angular momentum vector.
{\it Left panels:} randomly distributed spin orientations. {\it Right panels:} spin aligned along the orbital angular momentum vector. Each panel shows the 
PDF for mass ratios $q=0.1$ (\textit{solid line}), $q=0.25$ (\textit{dashed line}), $q=0.33$ (\textit{dotted line}) and $q=1$ (\textit{dash-dotted line}). 
For $\alpha_{1,2}=0$, $v_{\parallel}$ vanishes and the PDF is null.}  
\label{pdf}
\vskip 0.1cm
\end{figure} 

Figure~\ref{distributions} shows the distributions of $\vkick$, $v_\perp$, and $v_\parallel$ in the case where the black hole spins are either 
randomly oriented or aligned parallel to the the angular momentum vector. The maximum recoil velocity, $\sim 3,700\kms$, is attained in the direction parallel to 
$\vec{L}$ for maximally spinning black holes with anti-aligned spins oriented along the orbital plane, and $q=1$. The mean $\vkick$ exceeds $100\,\kms$ for mass ratios $q>0.1$. 
If spins are randomly oriented, then the maximum recoil velocity attainable in the orbital plane is smaller than the component along the orbital angular momentum. 
The maximum kick in the case $\alpha_{1,2}=0$ is of the order of $200\,\kms$ and is oriented in the orbital plane ($v_{\parallel}=0$).

Figure~\ref{pdf} shows the probability density distributions (PDF) of $\vkick$, $v_{\parallel}$, and $v_{\perp}$ at a given mass ratio, in cases where the spins are randomly 
oriented or aligned with the orbital angular momentum vector. The PDFs are calculated for mass ratios $q$=0.1, 0.33, 0.25, and 1. For randomly oriented spins, the peak of 
the $\vkick$ PDF shifts towards lower velocities for lower mass ratios, and the distribution broadens at higher mass ratios \citep[see also][]{lousto10}.

\section{Galaxy Merger Remnants}\label{hosts}

\begin{figure}
\centering
\includegraphics[width=0.4\textwidth]{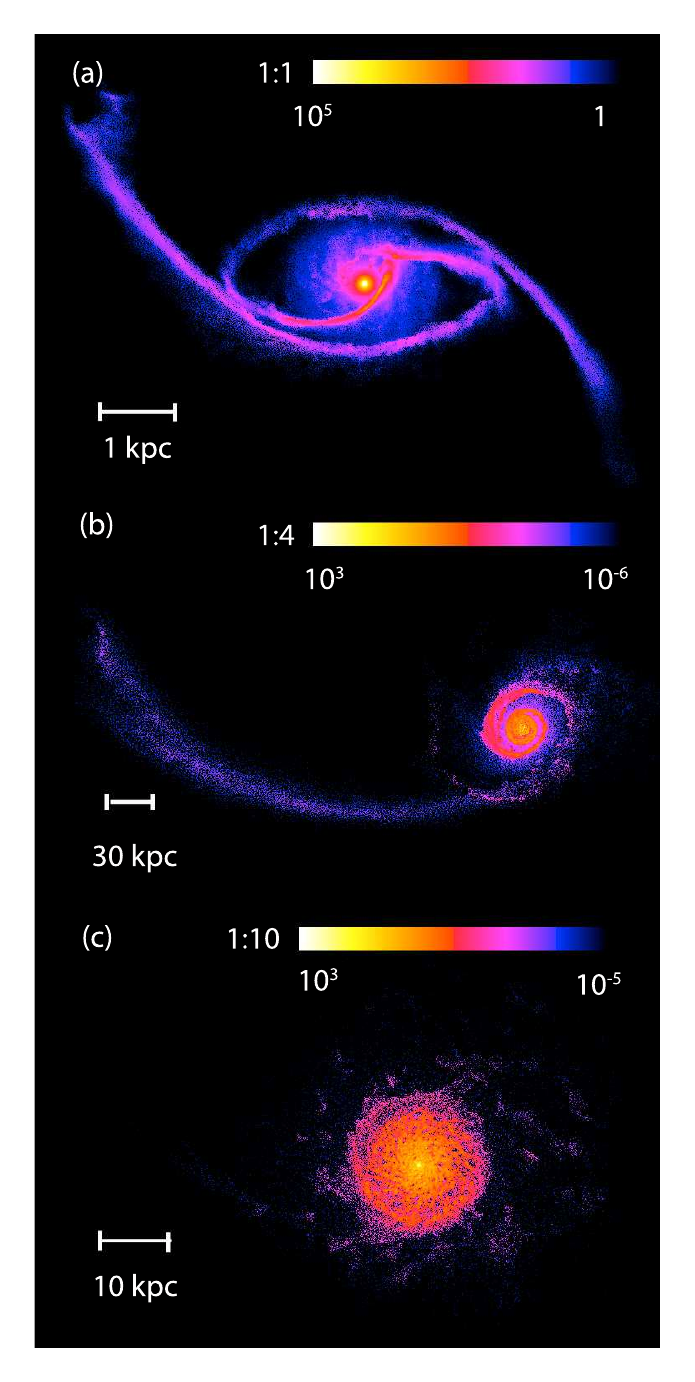}
\caption{\footnotesize Projected gas density of the galaxies hosting a recoiling MBH. {\it (a)} Remnant of a 1:1 merger simulation with initial gas fraction 
$f_g=0.1$ \citep{mayer07}. {\it (b)} Remnant of a 1:4 merger simulation with initial gas fraction $f_g=0.1$ \citep{callegari09}. {\it (c)} Remnant of a 1:10 
merger simulation with initial gas fraction $f_g=0.3$ \citep{callegari09}. The scaling is logarithmic and the highest and lowest values of the scale 
are given in units of atoms cm$^{-3}$. Note that the lower central density in the 1:4 merger remnant, which manifests as a shallower profile of the
inner rotation curve inside a few hundred parsecs relative to the 1:10 merger case, is an artifact of the larger gravitational softening adopted in this run.}
\label{host}
\end{figure}

We study the orbits of recoiling MBHs in three galaxy remnants that were previously simulated at high resolution, and in which the MBHs of the progenitor galaxies were shown
to successfully formed close pairs. The remnants are the endpoint of the 1:1 gas-rich major merger described in \cite{mayer07}, and of the 1:4 and 1:10 gas-rich minor
mergers of \cite{callegari09}.  Figures~\ref{host} and \ref{rotation} show a 2-D rendering of the hosts' projected density and the rotation curves of all 
components -- gas, stars, and dark matter  -- of the three remnants. 

\subsection{Major merger}\label{majormerger}
Our major merger remnant is part of the suit of simulations carried out \cite{mayer07}. 
These simulations tracked the formation of a MBH binary down to parsec scales following the collision between two spiral galaxies, and   
were run using the massively parallel $N$-body + smoothed particle hydrodynamics (SPH) code GASOLINE \citep{wadsley04}. Initially, two 
equal mass galaxies composed of a spherical dark matter halo, an exponential disk, a spherical stellar bulge, and a central MBH, were placed on a parabolic, 
coplanar, and prograde-prograde orbit. The galaxies have a virial mass of $10^{12} M_{\odot}$ and a baryonic disk mass of about $6 \times 10^{10} M_{\odot}$, with $10\%$ 
of the disk mass in a cold gas component ($T = 2 \times 10^4 K$) and the rest in stars (see \citealt{mayer07} for details on the structural parameters of the models).

Radiative cooling and star formation were implemented until the late stages of the simulation, when the dark matter halos had nearly merged and 
the baryonic cores were separated by about 6 kpc. At this point, in the reference run of \cite{mayer07}, the inner 30 kpc of the simulation volume was 
refined with the technique of particle splitting to achieve a spatial resolution of 2 pc. The star and dark matter particles were not splitted to limit the computational 
burden. The calculation that we use in this paper has a softening of 10 pc and a mass resolution in the gas component of $5000\,\msun$ after splitting 
(for a total of $N_g=1.4\times 10^6$ SPH particles). We use this run rather than the highest resolution simulation presented in \cite{mayer07} because 
this resolution matches those of the minor merger runs also used in this work, and because it allows to perform more recoil experiments at a reasonable computational 
cost (\citealt{mayer07} verified the convergence in the mass distribution of the merger remnants between the different resolutions).
The radiation physics in the refined region was modeled via an effective equation of state, an ideal gas with 
adiabatic index $\gamma=7/5$. Calculations that include radiative transfer have shown that this simple treatment approximates well the thermodynamics of a 
solar metallicity gas heated by a starburst over a wide range of densities \citep{spaans00, klessen07}. 

By the end of the simulation, $t=5$ Gyr after the merger, tidal tails continue to evolve in the outer parts of the remnant and over 60\% of the gas has been 
funneled to the central region, forming a compact nuclear gaseous disk of mass $M_d = 3 \times 10^9 \msun$, radial length $a = 75$ pc, and scale height $b = 20$ pc. 
The inner disk is turbulent and  rotationally supported. The two MBHs have sunk down from about 40 pc to a few parsecs in less than a million years, are 
gravitationally bound to each other, and their orbital decay is controlled by gas drag in the supersonic regime, not dynamical friction against the stellar background 
\citep{mayer07,escala04,ostriker99}. Figure~\ref{rotation} (left panel) shows the rotation curve for the different components of the galaxy remnant at the 
endpoint of the merger. The gaseous component dominates the dynamics of the inner 300 pc of the remant galaxy. The escape speeds from the center in the 
perpendicular ($\theta=90^\circ$) and parallel ($\theta=0^\circ$) directions to the angular momentum vector of the gaseous disk are $v_{e}^{\perp} = 
1,400\,\kms$ and $v_{e}^{\parallel} = 1,320\,\kms$, respectively. 

\subsection{Minor mergers}\label{minormerger}

The hierarchical $\Lambda$CDM structure formation scenario predicts a high frequency of minor galaxy mergers \citep{Lacey93,Fakhouri08}. \cite{callegari09} run a set of 
collisionless (``dry", with $f_g=0$) and gasdynamical (``wet") minor mergers with the same gas fraction in the primary and secondary galaxies, either $f_g=0.1$ or 0.3. 
As in the major mergers, the primary galaxy is a Milky Way-mass galaxy sampled with $10^6$ particles dark matter particles and (initially) 
$2\times10^5$ star and $10^5$ gas particles. The secondary galaxy is obtained by rescaling the primary galaxy in mass by 1/4 or 1/10 of its mass,
according to the scaling laws of \cite{mo98} for CDM galaxies, and is sampled by an equivalent number of particles as the primary. In the 1:10 merger, 
which is assumed to take place at $z=3$ since this is an ideal redshift window for LISA, the masses, radii and velocities
of the galaxies are scaled further by a factor $H(z)/H_0$ as expected in CDM \citep{callegari09}.

While no MBH pair was formed in the collisionless runs (the two holes were defined as a ``pair" if their relative orbit shrunk down to a separation equal to twice the softening, 
which sets the resolution limit), the presence of a  gaseous component allows the formation of a MBH pair because the massive gaseous cores in which the MBHs are embedded
sink efficiently by dynamical friction delivering the holes at the center of the remnant. Furthermore, the higher background density arising as inflows driven by
tidal torques associated with bar instabilities and shocks push gas to the center of the galaxies during the merger and enhance dynamical friction in the late stages of the orbital 
decay \citep{callegari09}. Here we consider two such dissipational runs: a 1:4 merger with initial gas fraction $f_g=0.1$, and a 1:10 merger with $f_g=0.3$. Both resulted in the 
formation of MBH pairs separated by twice the softening length (200 pc for the 1:4 merger and 40 pc for 1:10), and it is expected that the holes will subsequently merge 
on a timescale $<$ 1 Gyr \citep[see][]{callegari09}. The galaxies used in the simulation were initialized with a mixture of primordial metallicities. The runs were performed with the 
GASOLINE code, including this time radiative cooling, atomic line cooling (by H and He), and cooling via collisions, star formation, and blast-wave supernova 
feedback until the end of the simulation, rather than switching to an effective equation of state.  The resulting interstellar medium (ISM) density distribution 
is clumpier than in the case of the major mergers that adopt an effective equation of state and neglect supernova explosions. 

\begin{figure*}[th]
\centering
\includegraphics[width=.75\textwidth]{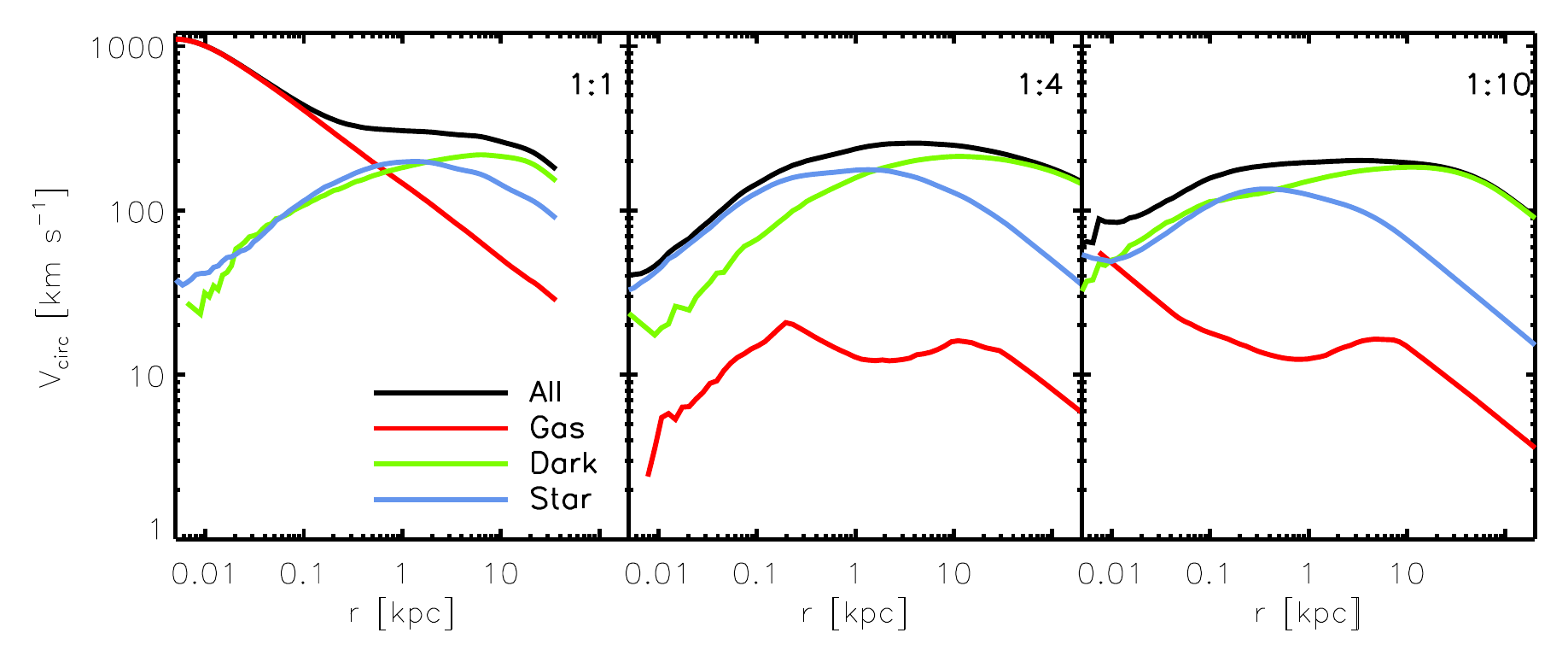}
\caption{\footnotesize Rotation curves of the galaxy remnants hosting a recoiling MBH. {\it Left panel:} 1:1 merger. {\it Middle panel:} 1:4 merger. {\it Right
panel:} 1:10. The colors represent the circular velocity profiles of the gas ({\it red}), stars ({\it blue}), dark matter (\textit{green}), and total (\textit{black}), 
respectively.} \label{rotation}
\vskip 0.3cm
\end{figure*}

The most important difference between the major and the minor mergers, however, is that bar-driven torques are much weaker in the latter case. This is especially true 
in the primary galaxy, which is only weakly destabilized by the secondary: as a result, gas inflows are much weaker, and $< 10\%$ of the gas
that was originally at large radii reaches the central hundred parsec region by the end of the merger, as opposed to more than $60 \%$ in the case of major 
mergers. Figure~\ref{rotation} (middle and right panels) shows the rotation curves of the minor merger remnants. They include gas that will eventually form stars; furthermore, 
heating and stirring by supernova feedback will make the gas more pressure-supported and extended.  Therefore, we expect the MBH return time in minor mergers to be largely 
determined by dynamical friction against the stars and the dark matter. In addition, the lower central densities imply a lower escape speed, so we expect that holes
with a given kick velocity will travel larger distances in this case compared to a major merger. The escape speeds from the center in the 
perpendicular ($\theta=90^\circ$) and parallel ($\theta=0^\circ$) directions to the angular momentum vector of the gaseous disk are $v_{e}^{\perp} = 
520\kms$ and $v_{e}^{\parallel} = 490\kms$ in the 1:4 merger, respectively, and $v_{e}^{\perp} = 490\,\kms$ and $v_{e}^{\parallel} = 450\kms$ in the 1:10 merger. 

\section{Simulations of Recoiling MBHs}\label{simulations}

We performed six simulations of recoiling MBHs in the 1:1 merger remnant of \cite{mayer07}. A MBH binary of total mass $M_{\bullet} = 5.2 \times 10^6 \msun$ was 
replaced with a single black hole of the same mass. The hole was given a kick of $\vkick$ = 400, 800, and 1,200 $\kms$ at an angle $\theta=0^{\circ}$ or $90^{\circ}$ relative 
to the angular momentum of the disk. In Section~\ref{kickvelocity} we showed that maximum recoil velocities are of order $200\,\kms$ if spins are aligned with the
orbital angular momentum of the binary. Here, we make no assumptions regarding the spin orientation or the orientation of the binary orbital plane relative to the 
gaseous disk, since the main objective of the simulations is to calibrate the semi-analytical model described in Section~\ref{model}. However, these configurations 
will enter in the calculation of the probability of observing recoiling MBHs as off-nuclear AGN during their wandering phase (see Section ~\ref{detectability}). 
For comparison, we have also run a simulation without the MBH. 

Our gas mass resolution is $5\times10^3 \msun$, for a total of $1.5\times10^6$ SPH particles, and the gravitational softening is 10 pc for the MBH 
as well as the SPH particles. The hole was kicked starting from where the binary was, at the center of mass of the gas distribution located some 20 pc 
away from the center of the potential. 
The simulations were run using GASOLINE, and were restarted using the same thermodynamical parameters as in the refined part of the \cite{mayer07} simulation, i.e. the SPH 
particles were treated as an ideal gas with adiabatic index $\gamma=7/5$.
Since only a small fraction of the nuclear gas turns into stars during the timescale of the simulations, we neglect star formation to reduce the computational burden.  
We run the simulations for up to 50 Myr, a compromise between obtaining a good calibration for our semi-analytical models and limit CPU time.  It has been 
shown by \cite{dotti07} that MBHs moving in nuclear disks with varying mass fraction of stars produce a drag (via dynamical friction) that is almost 
indistinguishable from the purely gaseous case. It is gas thermodynamics instead that determines the magnitude of frictional drag in both purely gaseous or 
star forming disks, as it sets the density and temperature distribution of the gaseous disk \citep{escala05,mayer07}. 
\begin{figure}[thb!]
\centering
\includegraphics[width=.31\textwidth]{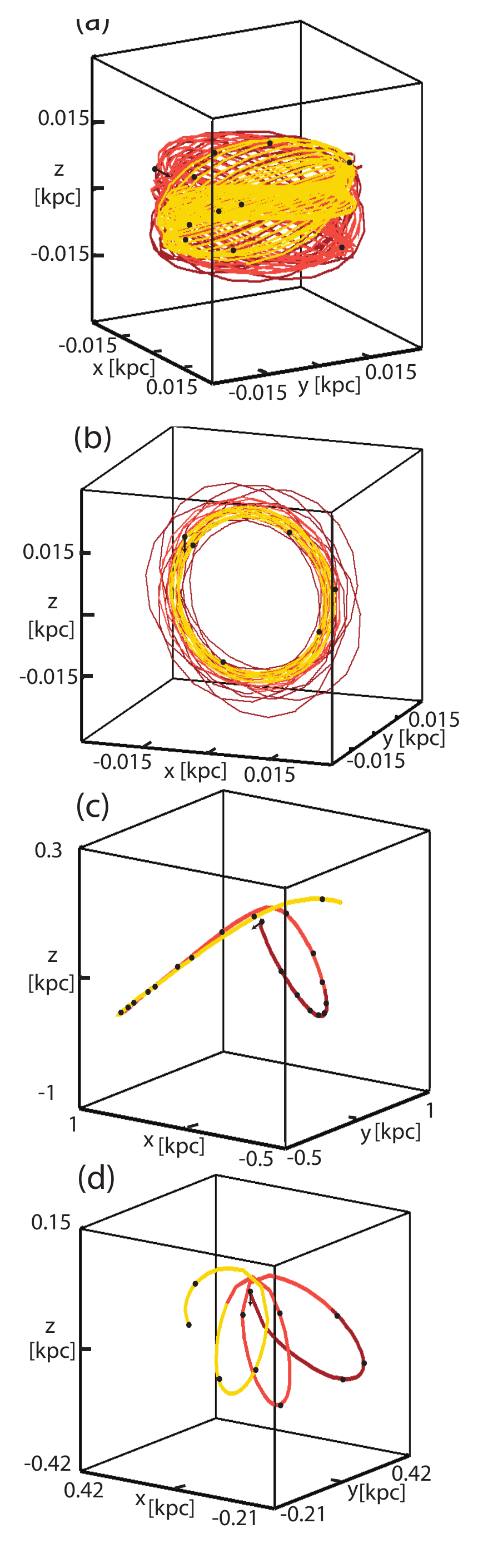}
\caption{Orbit of a MBH of mass $M_{\bullet} = 5.2\times10^6\msun$ in major merger remnant. The panels show a hole kicked with {\it (a)} 
$\vkick=800\kms$ and $\theta=90^{\circ}$ (parallel to the gaseous disk plane), {\it (b)} $\vkick=800\kms$ and $\theta=0^{\circ}$, {\it (c)} $\vkick=1,200\,\kms$ 
and $\theta=90^{\circ}$, and {\it (d)} 
$\vkick=1,200\,\kms$ and $\theta=0^{\circ}$. The colors indicate time along the orbit, from $t< t_{\rm run}/3$ (\textit{brown}) to $t>2t_{\rm run}/3$ (\textit{yellow}). 
The runtime of the simulation is $t_{\rm run}=10$ Myr for $\vkick=800\kms$ and $t_{\rm run}=20$ Myr for $\vkick=1,200\,\kms$. The black dots mark 1 Myr intervals. 
Orbits are plotted with respect to the deepest point in the disk potential.}  
\label{orbits}
\end{figure}
%

%
\begin{figure*}[thb!]
\centering
\includegraphics[width=0.8\textwidth]{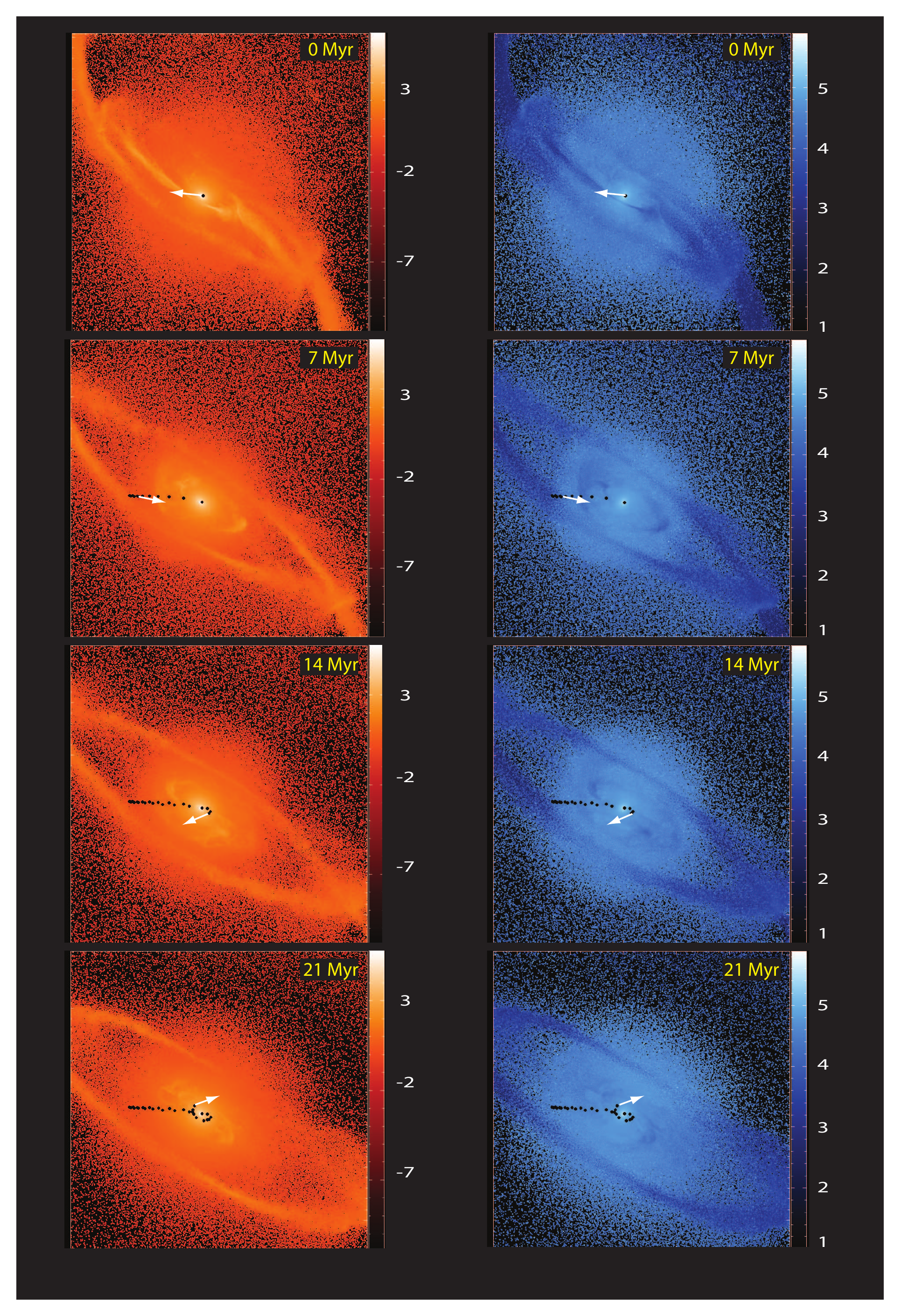}
\caption{The orbit of a recoiling MBH with $M_\bullet=5.2\times10^6\msun$, $\vkick=1,200\,\kms$, and $\theta=90^{\circ}$ (\textit{black dots}), as in panel 
{\it (c)} of Figure 6. The white arrows show the instantaneous direction of the velocity vector at the time of the corresponding snapshot. The color 
coding indicates gas number density in units of $\log (n/{\rm cm}^{-3})$ ({\it left panel}) and specific internal energy in units of $\log (u/\kmss)$ ({\it right panel}).}
\label{orbit}
\end{figure*}
%

%
\begin{figure*}[thb]
\centering
\includegraphics[scale=0.36]{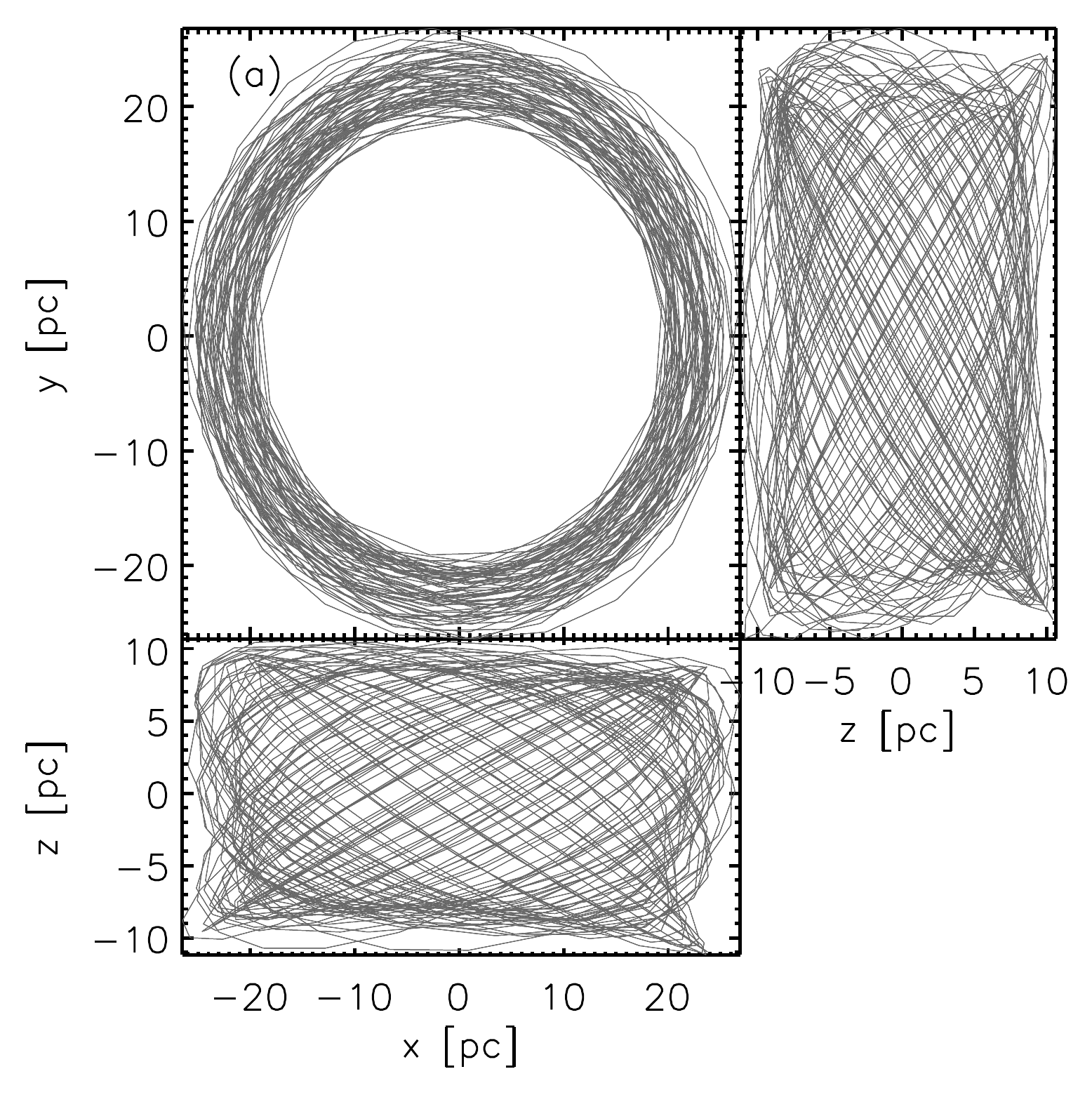}
\includegraphics[scale=0.36]{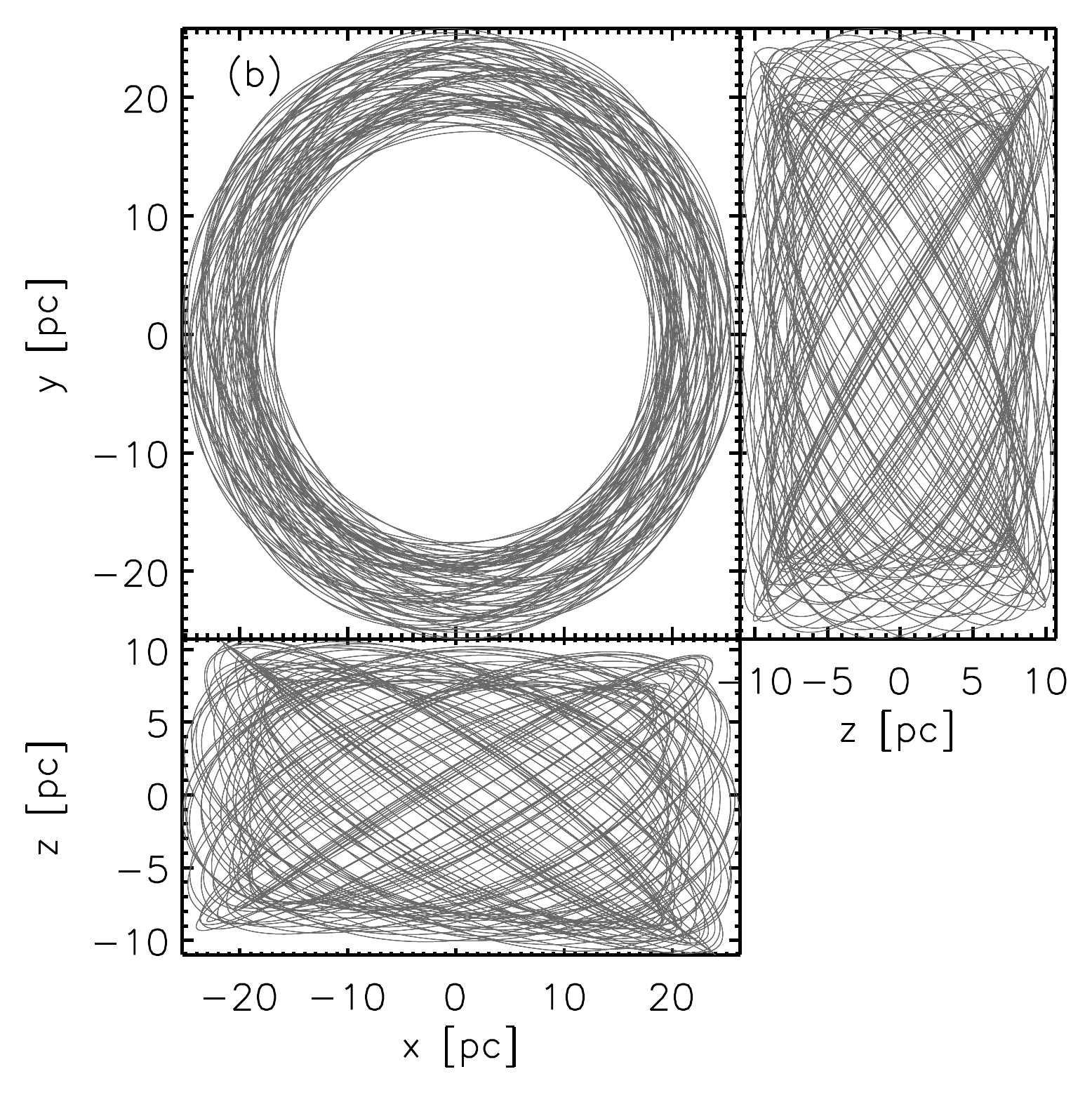}
\includegraphics[scale=0.45]{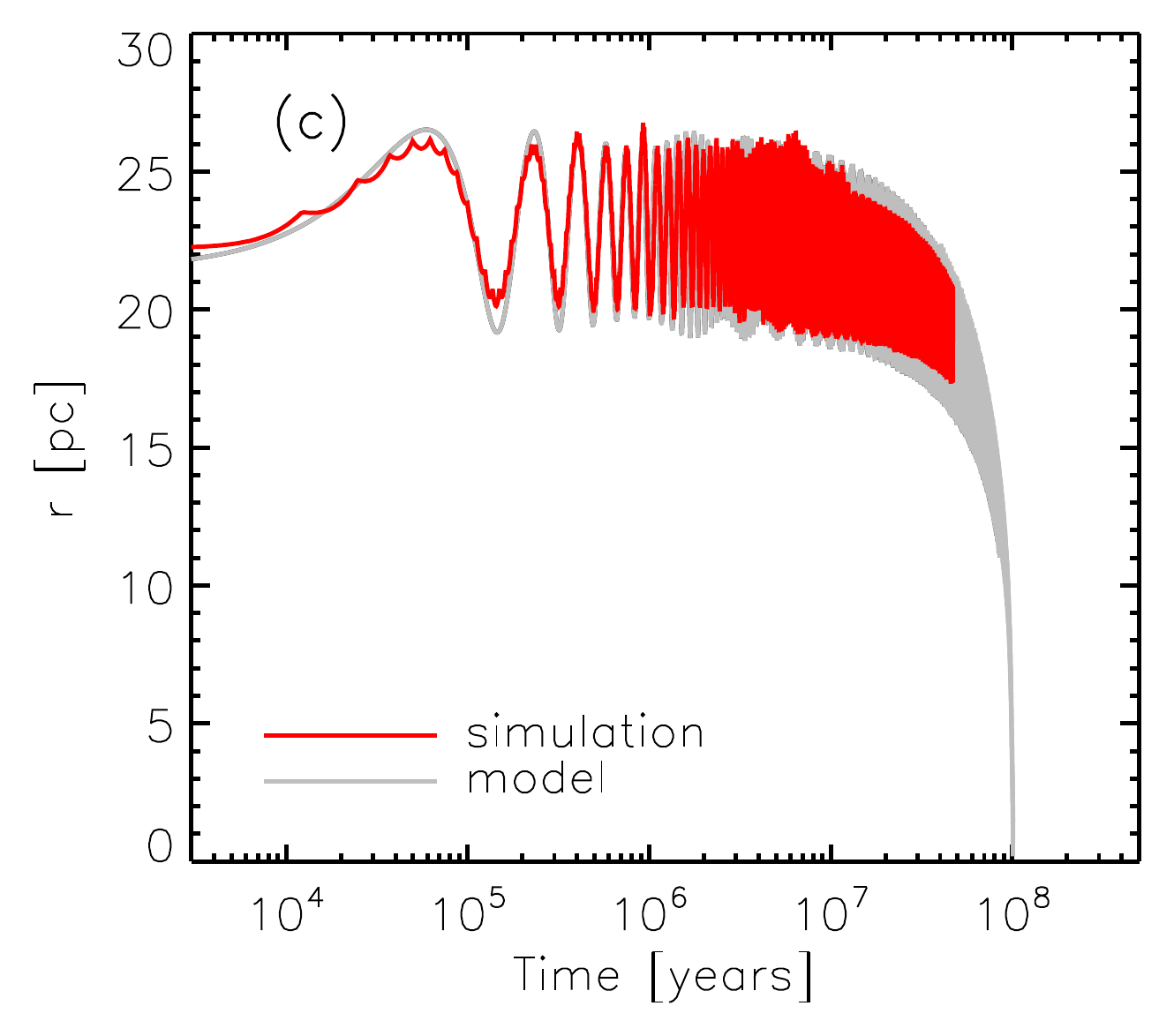}
\caption{Comparison between the orbit of a recoling MBH of mass $M_\bullet=5.2\times 10^6 \msun$ in our numerical simulation versus the 
semi-analytical model. {\it (a)} Same simulation as in panel {\it (a)} of Figure 3. {\it (b)} Semi-analytical orbit. For clarity, panels {\it (a)} and 
{\it (b)} show the orbit 10 Myr after the initial kick. {\it (c)} Radial evolution of the orbit in the semi-analytical model ({\it gray}) compared to the simulation ({\it red}).}
\label{model_comp}
\vskip 0.3cm
\end{figure*}

It is important to emphasize that, in a model with an effective equation of state, the temperature should not be associated with the physical temperature of a specific 
phase of the interstellar medium in the nuclear disk, e.g. the molecular gas that makes up most of the disk mass \citep{Downes98}, but rather with an 
effective pressure that provides a realistic match to the pressure scale height of the disk. In a realistic case the cold molecular phase is embedded in a 
hot ionized medium that fills most of the volume and sets the actual pressure scale height. The pressure scale height of our nuclear disk is $z_0=$15 pc at 
the beginning of the simulations, and $z_0=$21 pc at $t=$20 Myr, and compares well with the pressure scale height of 23 pc observed in Mrk 231 \citep{Downes98,Davies04}. 
Likewise, the effective sound speed used throughout the paper (calculated directly from the simulations) should be close to the sound speed of the hot, diffuse phase 
(the cold molecular phase has a sound speed of $< 1\kms$). The peak of the sound speed in the major merger is $200\,\kms$, while it is $34\,\kms$ and $14\,\kms$
in the 1:4 and 1:10 mergers, respectively.

Figure~\ref{orbits} shows the resulting orbits for MBHs kicked with $\vkick=800\kms$ and $\vkick=1,200\,\kms$ in a direction that is either along the 
disk plane (cases {\it a} and {\it c}) or perpendicular to it (cases {\it b} and {\it d}). The boxes have varying lengths and the orbits are centered on the point where the potential 
is the deepest. The orbit for the case $\vkick=1,200\,\kms$ and $\theta=90^{\circ}$ is shown again in Figure \ref{orbit}, together with the evolution of the background gas number density 
and specific internal energy. 

\section{Semi-Analytical Model}\label{model}

To obtain a large sample of MBH orbits at relatively low computational cost, we have constructed a semi-analytical model based on our numerical results. 
We can then follow the trajectories of MBHs until they return to the center of their host, on timescales that can be much longer than the time it is practical 
to run a full $N$-body + SPH simulation, particularly for larger kick velocities. The motion of a MBH in a multi-component potential can be studied 
analytically by integrating numerically its equation of motion under the action of a conservative force $\vec{\nabla}\Phi$ and a damping frictional term:
\begin{equation}\label{eqm}
\frac{d\vec{v}}{dt} = -\vec{\nabla}\Phi + \vec{ f}_\mathrm{DF},
\end{equation}
where the total potential is $\Phi = \Phi_{\rm gas} + \Phi_{\rm stars} + \Phi_{\rm dark}$. The gradient of the potential is computed directly from the simulation data. 
The dynamical friction term for an object of mass $M_\bullet$ moving with relative velocity $v$ through a gaseous medium of density $\rho$ is given by
\begin{equation}\label{df}
\vec{f}_{\rm DF} = - \frac{4\pi G^2 M_{\bullet}\ln \Lambda \rho}{v^3}\,  I_v\, \vec{v}.
\end{equation}
The velocity integral, $I_v = I(\mathcal{M})$, depends on the Mach number $\mathcal{M} \equiv v/c_s$, where $c_s = \sqrt{\gamma k_B T / \mu}$ is 
the sound speed. In our model, the local gas density and sound speed are measured directly from the simulation data in a sphere of radius 
\begin{equation}
R_{\rm sph}(r)=R_{\rm sph}(1\,{\rm kpc})\left[{\rho(1\,{\rm kpc}) \over \rho(r)}\right]^{1/3} 
\end{equation}
around the MBH, where $\rho(r)$ is the average density at a distance $r$ from the host center and we choose a reference radius $R_{\rm sph}(1\,{\rm kpc})=100\,$pc.
As for the shape of the velocity integral, we adopt the hybrid formulation of \cite{tanaka09}, who combined the prescriptions of \cite{ostriker99} and \cite{escala04} 
into:

\begin{eqnarray}
I(\mathcal{M}) = \left\{
    \begin{array}{l l}   
         0.5 \ln \Lambda \left [ {\rm erf} \left( \frac{\mathcal{M}}{\sqrt{2}}\right )-\sqrt{\frac{2}{\pi}} \mathcal{M} \exp\left ({-\frac{\mathcal{M}^2}{2}}\right ) \right ] \\
        (0\leq \mathcal{M} \leq 0.8);\\
        1.5 \ln \Lambda \left [ {\rm erf} \left( \frac{\mathcal{M}}{\sqrt{2}}\right )-\sqrt{\frac{2}{\pi}} \mathcal{M} \exp\left ({-\frac{\mathcal{M}^2}{2}}\right ) \right ] \\
         (0.8<\mathcal{M} \leq 1.5);\\
        0.5\ln \left ( 1-\frac{1}{\mathcal{M}^2}\right ) + \ln\Lambda \\
         (\mathcal{M} > 1.5). \\
    \end{array}\right.
\end{eqnarray}
The Coulomb logarithm is set to $\ln\Lambda=3.1$ \cite{escala04}.
In a collisionless medium, the velocity integral depends on the local velocity distribution of the stars or dark matter. If the medium is isotropic and the velocity 
distribution is Maxwellian, then $I_v = I(X\equiv v/\sqrt{2}\sigma)$, where $\sigma$ is the one-dimensional isotropic velocity dispersion. 
Equation (\ref{df}) then becomes the Chandrasekhar dynamical friction formula \citep{chandrasekhar43}, where
\begin{equation}
I(X) = {\rm erf} (X)-\frac{2X}{\sqrt{\pi}} e^{-X^2} 
\end{equation}
and $\rho$ is the stellar or dark matter density. In a recent merger, however, the velocity distribution of the stellar and dark component will 
deviate from Maxwellian. We therefore calculate the velocity integral numerically from the actual velocity space distribution of the particles in the simulations,
\begin{equation}
I_v = \int_0^v f(v')dv'
\end{equation}
where $f(v')dv'$ is the number of particles with velocities between $v'$ and $v'+dv'$ in the vicinity of the hole. The equation of motion of the MBH is integrated 
numerically using an adaptive Adams-Bashforth-Moulton algorithm.

\begin{figure}[hb]
\centering
\includegraphics[scale=0.4]{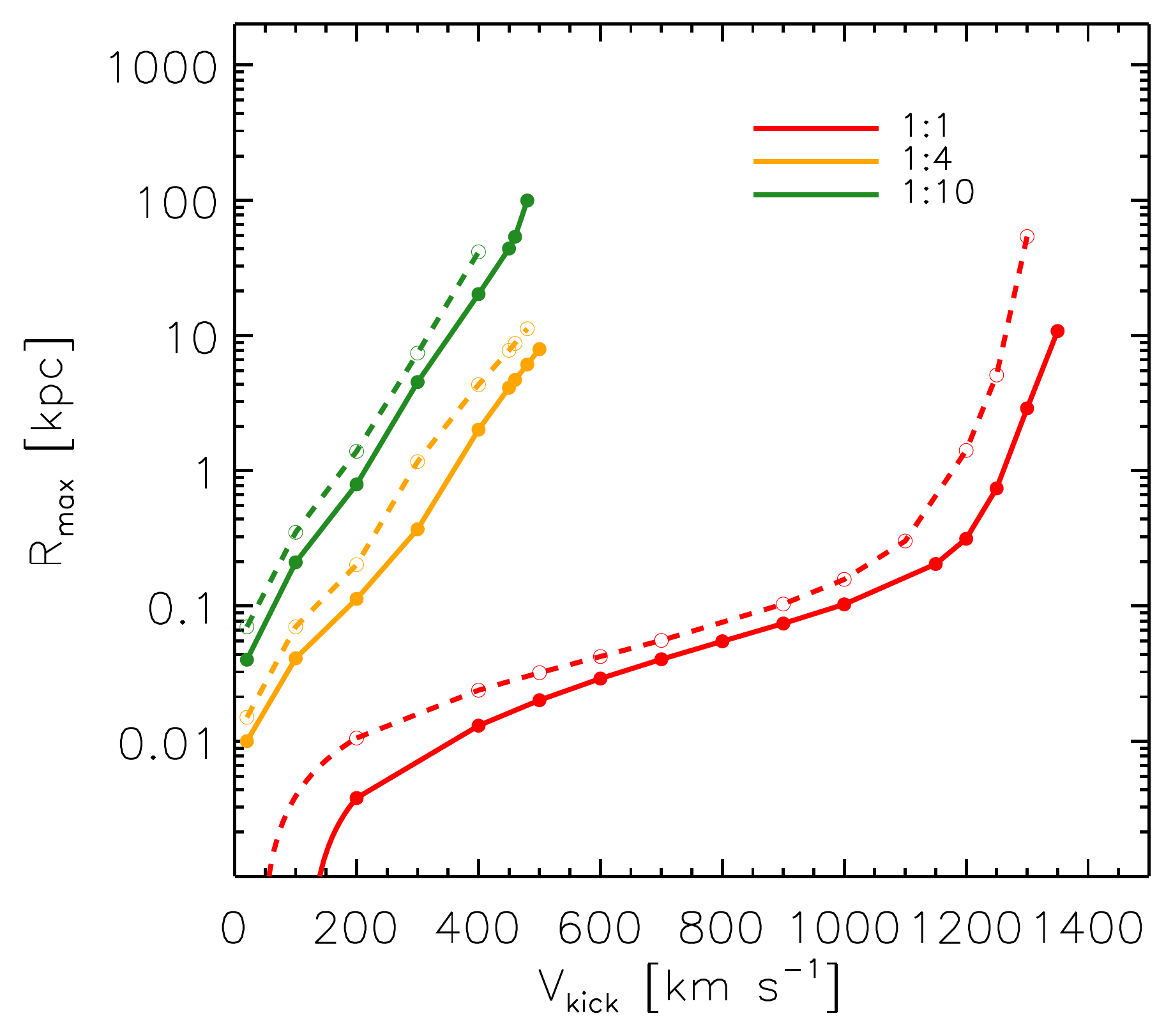}
\caption{First apocenter distance of recoiling MBHs as a function of kick velocity. \textit{Red:} 1:1 merger remnant. \textit{Orange:} 1:4. \textit{Green:} 1:10. 
\textit{Solid curves:} MBHs recoiling in a direction perpendicular to the disk angular momentum vector. 
\textit{Dashed curves:} MBHs recoiling in a direction parallel to the disk angular momentum vector.
}
\label{apocenter}
\vskip 0.3cm
\end{figure}

Figure ~\ref{model_comp} shows a comparison between the orbits of a MBH in our numerical simulations versus the semi-analytical model. The hole was kicked 
with $\vkick=800\,\kms$ in the gaseous disk of the major merger remnant. The trajectory is shown in projection in the left and middle panels, while
the right panel depicts the radial evolution of the orbit. Over the timescale covered by the simulation, the semi-analytical model is clearly able to reproduce 
both the amplitude of the oscillations (dictated by the potential) and the decay by friction. Figure~\ref{apocenter} shows the first apocenter distance 
$R_{\rm max}$, estimated from our semi-analytical approach, for MBHs recoiling in our 1:1, 1:4, and 1:10 merger remnants along a direction that is either parallel or 
perpendicular to the disk angular momentum vector. The black hole mass was scaled in each case to match the $M_\bullet-\sigma$ relation, yielding 
$M_{\bullet} = \{5.2, 3.2, 2.8\}\times 10^6\,\msun$ for the 1:1, 1:4, 1:10 mergers, respectively. At supersonic speeds, gas drag plays an important role in damping the 
MBH orbit in the 1:1 remnant. Shortly after the kick, when the hole's is moving at Mach number $1 < \mathcal{M} < 5$, gas drag accounts for over 70\% of the total frictional force. 
In the deep potential well of the remnant, the gas density can reach $10^5$ particles cm$^{-3}$, and MBHs kicked in the plane of the disk with initial velocities as 
high as $1,200\,\kms$ reach apocenter distances of only 0.3 kpc. As $\vkick$ approaches the escape speed along this direction, 
i.e. $\vkick=1,350\,\kms=0.98 v_e^{\perp}$, the first apocenter distance reaches $R_{\rm max}= 10$ kpc. 

MBHs recoiling in the direction parallel to the disk angular momentum vector can reach the low-density outer regions for slightly lower initial kicks, and therefore reach larger
apocenter distances. A hole with $\vkick=1,200\,\kms = 0.91v_e^{\parallel}$ reaches $R_{\rm max} \simeq 1$ kpc and has a return timescale that is close to 10 Gyr.  
Minor mergers have significantly lower central density profiles, and apocenter distances can be significantly larger even for modest recoil speeds. 
In the $1:4$ merger remnant, a kick of $\vkick=500\,\kms$ perpendicular to the disk can displace the hole by $R_{\rm max} \sim 10$ kpc. The same initial 
speed yields a maximal separation of $R_{\rm max}=100$ kpc in the $1:10$ case.

\section{Detectability of recoiling MBHs}\label{detectability}

Three main methods have been proposed to detect a recoiling hole: 1) through the imprint of the moving MBH on the gaseous medium of the host galaxy; 2) 
as a kinematically-offset AGN, as the broad line region (BLR) carried by the hole is displaced by a velocity $\Delta v$ from the narrow starlight associated 
with the host galaxy; 3) as a spatially-offset AGN, where the accreting MBH is observed to be displaced by a distance $\Delta r$ from the host's nucleus. 

\subsection{Interaction with the gaseous disk}\label{imprint}

\cite{devecchi09} recently carried out a suite of numerical simulations aimed at studying the detectability of recoiling MBHs in the hot gas 
of elliptical galaxies. The host galaxy was modeled as a spherical Hernquist sphere \citep{Hernquist90}, with a mass resolution of $2\times 10^7\msun$
and a polytropic equation of state with $\gamma=5/3$. They found that, at high Mach numbers, the shocked gas can produce an X-ray signature that may
be detectable with the \textit{Chandra} satellite in nearby galaxies. Simulations of recoiling holes surrounded by a thin circumbinary disk 
\citep{rossi10} show that the likelihood of observable emission is low and depends strongly on the orientation of the kick relative to the thin disk, 
on the magnitude of the kick, and on the mass of the disk. 

In the numerical simulations described in Section~\ref{simulations}, the inner disk is rotationally supported ($v_{\rm rot} = 
300\kms$), turbulent ($v_{\rm turb}=500\kms$ at $r = 20$ pc), and actively accreting material 
from the outer regions. Dense columns of cooler gas reach the nuclear region, constantly changing its internal energy profile. 
We find that the ratio between the integrated energy deposition of the MBH (associated with energy losses from gas frictional drag) 
and the integrated change in internal energy by the gas (measured in the simulation without MBH) is very small, $\Delta E_{\rm DF}/\Delta E_{\rm int}=0.02$. 
As a result, overdensities produced in the gas by the recoiling MBH are rapidly weakened. In contrast with the results of \cite{devecchi09}, the removal 
of the central hole from the disk potential does not have a significant effect on the central density profile, since the mass of the hole amounts to only 
$2\%$ of the mass of the nuclear disk. At the end of the major merger $+$ MBH simulation, the gas density and internal energy profiles (Fig. ~\ref{orbit}) 
are nearly undistinguishable from those in the no-MBH run. 

By contrast, in the 1:10 merger, the integrated change in gas internal energy is comparable to the orbital energy lost by the MBH over a few tens of orbits. 
High Mach numbers can be reached at lower recoil speeds since the sound speed profile rises slowly to reach $30\,\kms$ at 50 pc and then stays constant 
out to 100 pc (whereas in the major merger the sound speed reaches a peak of $230\,\kms$ at 10 pc and then drops to $150\,\kms$  at 100 pc).
This suggests that minor merger remnants may be suitable sites for the detection of recoiling MBHs through their interaction with ambient material.

\subsection{Off-nuclear AGNs}\label{offset}

The detectability of MBHs as offset AGNs depends on the hole's phase-space location at the time of observation and on its ability to accrete material and shine
at that location. We assume that, upon ejection, the hole carries a punctured accretion disk \citep[e.g.][]{loeb07} and an associated BLR that is
visible for a time comparable to the viscous lifetime of the disk. The mass carried by the disk is \citep{loeb07,volonteri08}
\begin{equation}\label{mdisk}
M_{\rm disk} = 1.9\times 10^6 \alpha_{-1}^{-0.8} \zeta^{-0.6}M_7^{2.2} V_3^{-2.8}\msun, 
\end{equation}
where $\alpha_{-1} $ is the viscosity parameter in units of the fiducial value 
$\alpha=0.1$, $M_7$ is the MBH mass in units of $10^7M_{\sun}$, $\zeta \equiv \epsilon_{-1}/f_E$, $\epsilon_{-1}$ is the matter-to-radiation conversion efficiency 
in units of 10\%, $f_E \equiv \dot M/\dot M_E$ is the hole's accretion rate in units of the Eddington rate $\dot M_E \equiv 4 \pi G M_{\bullet} m_p/(\epsilon 
\sigma_Tc)$ (we use $f_E=0.1$), and  $V_3$ is the kick velocity in units of $10^3\,\kms$. The condition $M_{\rm disk}<M_\bullet$ requires 
\begin{equation}
\vkick> 550 \, \alpha_{-1}^{-2.8} \zeta^{-0.21}M_7^{0.43}\,\kms.
\end{equation}
When the above equation is not satisfied, we impose $M_{\rm disk}=M_\bullet$: this yields an AGN lifetime equal to $t_{\rm AGN}\equiv M_\bullet/\dot M=\epsilon 
c \sigma_T/(4\pi G m_p f_E)\approx 45\,\zeta \, {\rm Myr}$. The condition $M_{\rm disk} < M_{\bullet}$ also sets the maximum outer radius of the disk
\begin{equation}
R_{\rm out}< 150\,\alpha_{-1}^{-0.57} \eta^{-0.43}M_7^{-0.86} R_s,
\end{equation}
where $R_s = 2GM_{\bullet}/c^2$ is the Schwarzschild radius of the hole. At large radii ($> 0.1$ pc), gas in the disk may be gravitationally unstable and prone to star 
formation \citep{collin99}. However, the maximum disk size considered in our model is $\sim 10^{-3}$ pc, and therefore we neglect this effect in our analysis. 
We also find that ram pressure stripping will not affect the accretion disk as the MBH travels through the host. For a hole moving with velocity $v$ relative to a 
medium of density $\rho$, we can calculate the stripping radius $R_{\rm str}$ by equating the ram pressure force $F_{\rm ram}=\rho v^2$ to the 
restoring force of the disk, using the density profile of \citet{goodman04}. In the cases considered here, $R_{\rm out} \ll R_{\rm str}$, and therefore the disk is able
to survive.

Because precise measurements of the peaks of broad emission lines in quasar spectra are challenging, the observation of potential recoil candidates as kinemtically-offset 
AGNs -- whose broad emission lines are substantially shifted in velocity relative to the narrow-line gas left behind \citep{bonning07} -- 
is biased toward large kicks. In the following we shall assume that a recoiling hole is detectable as a kinematically-offset AGN if its velocity shift, $\Delta v$, is above 
the detection threshold $\Delta v_t = 600\,\kms$. At every time step of our numerical integration of the hole's orbit, we 
calculate the instantaneous accretion rate, the associated luminosity, and update the MBH mass. When the initial bound disk is exhausted, the BLR is assumed to 
vanish and the recoiling MBH is no longer recognizable as such from kinematics measurements. From this point on, the MBH is allowed to accrete material from the ISM. 
The fueling rate depends on the local density and sound speed of the gaseous medium and on the speed of the hole following
the Bondi-Hoyle formula:
\begin{equation}
\dot{M}_B = 7\times10^{-7}\ n_0 M_7^2 v_3^{-3} \msuny,
\end{equation}
where $n_0$ is the local gas number density of the ISM in units of cm$^{-3}$, $v_3 \equiv (v^2+c_s^2)^{1/2}$ is expressed in units of $10^3\,\kms$,  and $v$ is the instantaneous 
MBH velocity relative to the host. The mass accretion rate in this phase is given by the minimum between the Bondi rate and the Eddington rate, and the bolometric luminosity is then
\begin{equation}
L = \epsilon c^2\times \min(\dot{M}_B,\dot{M}_E).
\end{equation}
To be detected as a spatially-offset AGN, a recoiling accreting hole must be both above a threshold spatial separation $\Delta r_t$ from the host center 
as well as above a luminosity threshold $L_t$. We set $\Delta r_t$ equal to roughly twice the resolving power of the telescope -- 0.2" for the {\it Hubble Space Telescope
(HST)} and the {\it James Webb Space Telescope (JWST)}, and 1" for {\it Chandra}. For reference, at redshift $z=1$ (0.1) an offset of 10 kpc corresponds to an angular separation 
of 1.25" (5.4"). In the following, we set the threshold X-ray luminosity to $L_{t,X}  = f_{\rm lim} 4\pi D_L^2$ where $f_{\rm lim}=3.8\times10^{-16}\, \rm erg\,cm^{-2}\,s^{-1}$ is the on the on-axis \textit{Chandra} limit and $D_L$ is the luminosity distance, and use a bolometric to X-ray correction of $\chi_{2-10\,\rm keV}= 20$ if $L/L_E <0.1$ 
and $\chi_{2-10\,\rm keV}= 50$ otherwise \citep{vasudevan2007}. The optical B-band luminosity threshold $L_{t,B}$ is calculated using a limiting flux of $f_{\rm lim}=2\times10^{-14}, \rm erg\,cm^{-2}\,s^{-1}$ from observed broad-line AGN \citep{zakamska03} and the B-band bolometric correction is taken from \citet{marconi04}. Both X-ray and optical thresholds 
are consistent with observational limits from multi-wavelength studies of AGN \citep[e.g][]{pierce10}. 
%
\begin{figure}[th]
\centering
\includegraphics[width=.5\textwidth]{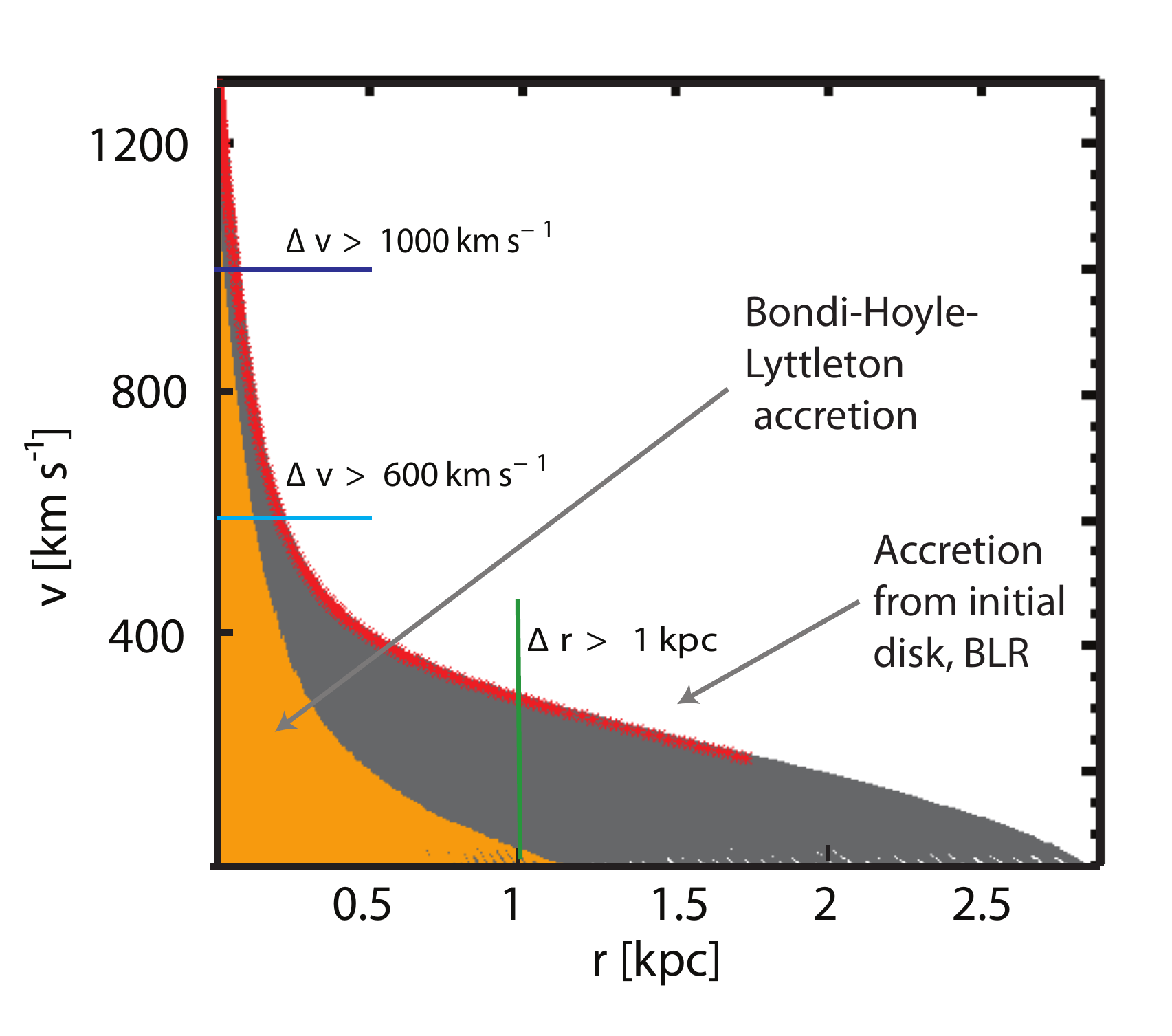}
\caption{Phase diagram of the orbit of a MBH of mass $M_{\bullet} = 5.2\times10^6\msun$ recoiling with 
$\vkick=1,300\,\kms$ along the nuclear disk of the major merger remnant at $z=0.025$. The MBH undergoes disk accretion and is shining 
with an X-ray luminosity $>L_{t,X}$ along the thin red line, and is powered by ISM accretion in the orange region.
}
\label{phase}
\vskip 0.3cm
\end{figure}

Figure~\ref{phase} shows a breakdown of the orbital phase diagram of a MBH of mass $M_\bullet = 5.2\times10^6\,\msun$ recoiling with velocity 
$\vkick=1,300\,\kms$ in the plane of the disk of the major merger remnant at $z=0.025$. Fiducial regions where $\Delta v > 600\,\kms$, 
$\Delta v >  1,000\,\kms$, and $\Delta r> 2"$ are marked in the velocity-distance plane. In this example $2"$ correspond to 1 kpc. The MBH 
undergoes disk accretion and is shining with an X-ray luminosity above $L_{t,X}$ along the thin red line: 
it is therefore detectable as a kinematically offset AGN when it is still close to the center, along the portion of the red line where $\Delta v \ge 600\kms$.
The MBH is instead detectable as a spatially-offset AGN in the low-velocity region where it accretes rapidly from the ISM (orange) or from the initial disk (red) 
and where $\Delta r> 1\,$kpc at this redshift. Note that large kinematic offsets occur near pericenter and can be observable for a shorter timescale
than large spatial offsets, which occur near apocenter.
In the following, we study in details the detectability of recoiling MBHs in two cases. \\

%
\begin{figure}[hb]
\centering
\includegraphics[width=.48\textwidth]{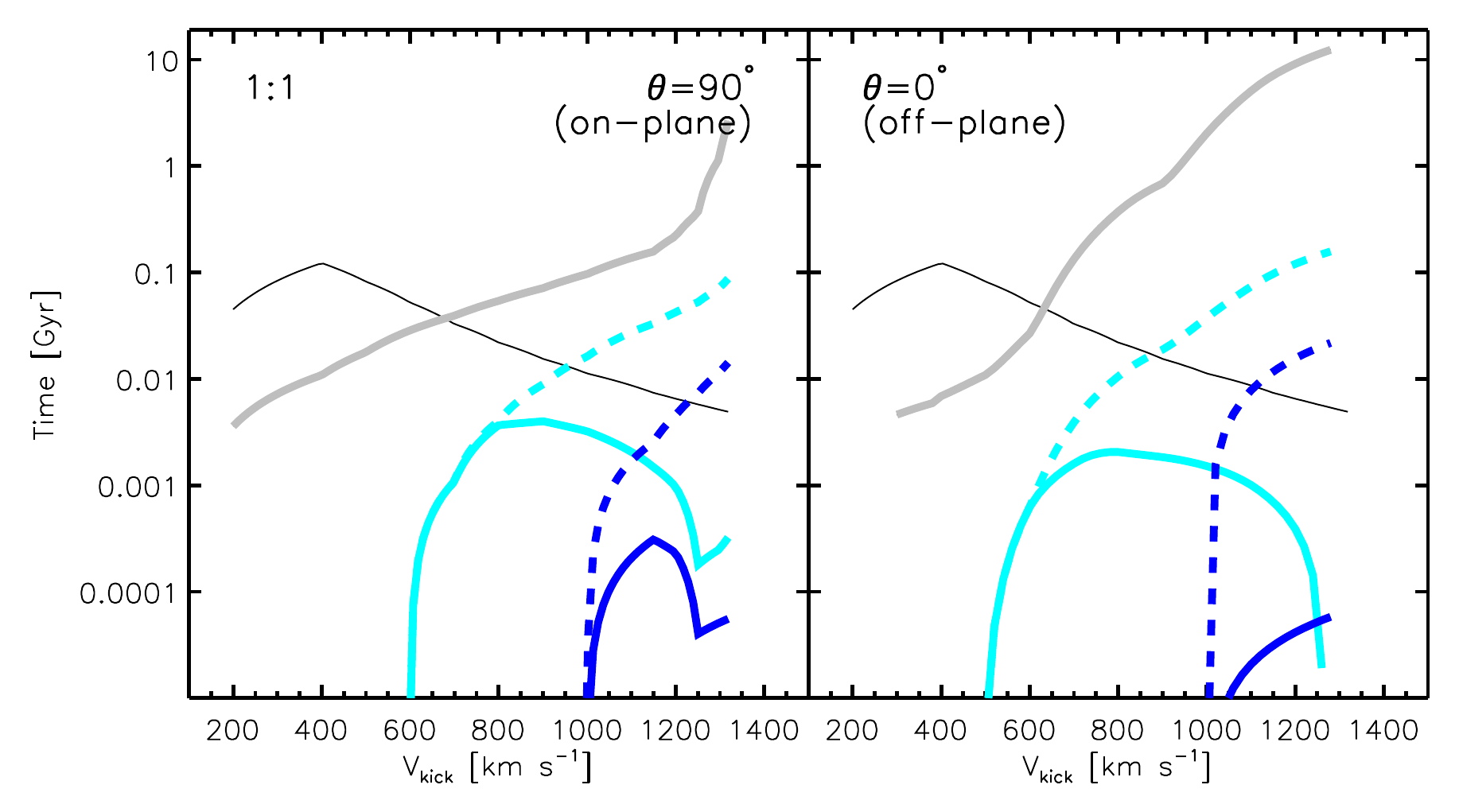}
\caption{Relevant timescales for assessing the detectability of recoiling MBHs as kinematically-offset AGNs in the major merger remnant, as a function of the 
kick velocity. \textit{Gray curves}: return timescale of a MBH of mass $M_{\bullet} = 5.2\times10^6\msun$. \textit{Dashed cyan curves}: total time during 
which the MBH has velocity offset $\Delta v \ge 600\kms $, i.e. 
above the threshold for the detectability of kinematically offset AGNs. \textit{Solid cyan curves}: total time over which the MBH has a velocity offset 
$\Delta v \ge 600\kms $ while shining with an optical luminosity $\ge L_{t,B}$ (for $z=0.1$). The dashed blue and solid blue curves use a threshold for detection of kinematically 
offset AGNs of $\Delta v \ge 1,000\,\kms$. \textit{Solid black curve}: lifetime of the accretion disk carried by the hole as a function of recoil velocity.}
\label{timescales_dv}
\vskip 0.3cm
\end{figure}

\noindent \underline{Case 1}. A best-case scenario for the detectability of recoiling MBHs as offset AGNs, in which: (a) the spins are randomly oriented 
with respect to the orbital angular momentum prior to the merger of the MBH binary (a possible mechanism for misalignment of the spins is chaotic accretion, see  \citealt{king06});
(b) the orbital plane of the binary is randomly oriented with respect to the plane of the gaseous disk (e.g. because of turbulent motions at sub-pc scales that may be 
uncorrelated with the angular momentum of the gaseous disk); and (c) 1:1, 1:4, and 1:10 galaxy merger remnants host MBH binaries of mass ratios $q=1, 0.25, 0.33$, respectively. 
In a recent study, \cite{callegari10} showed that during a 1:10 merger the primary galaxy may induce strong tidal torques onto the satellite that can result in the 
rapid growth of the smaller black hole, yielding in some cases binary mass ratios of $q=0.5$. In simulations where the gas fraction was set to $f_g=0.3$ and the merger was 
coplanar, a setup similar to the our initial 1:10 merger, the final mass ratio of the binary was found to be $q=0.33$.\\

\noindent \underline{Case 2}. A worse-case scenario for the detectability of recoiling MBHs as offset AGN, in which: (a) 
the spins are aligned with the orbital angular momentum vector of the binary prior to the merger event. Studies of the spin evolution of MBHs in a gaseous medium show that 
spins tend to align in gaseous environments due to frame dragging, yielding maximum recoil velocities of $\sim 200\,\kms$ \citep{bogdanovic07}. This effect 
drastically decreases the probability of observing both kinematically and spatially offset AGNs in gaseous environments; (b)
the orbital plane of the binary is aligned with the gaseous disk. This assumption, together with (a), implies that all recoils occur in the plane of the  disk;
and (c) 1:1, 1:4, and 1:10 galaxy merger remnants host MBH binaries of mass ratios $q=1, 0.25, 0.1$.

In both cases we consider a random uniform distribution of spin magnitudes in the range $0 \le a_{1,2} < 1$. 
Conservation of linear momentum also requires that the mass of the disk carried by the hole reduces the initial kick velocity to an effective value
\begin{equation}\label{veff}
V_{\rm eff} = \vkick \left ( \frac{M_{\bullet}}{M_{\bullet}+M_{\rm disk}} \right). 
\end{equation}
We account for this effect when estimating the detection probabilities associated with recoiling MBHs below.

\subsection{Kinematically-offset AGNs}\label{dv}

Several candidates of velocity-offset AGNs have been discussed in the literature \citep[e.g.][]{Magain05,Komossa08,shields09,civano10}, but none
has been positively confirmed to date. 

The study of \cite{bonning07} shows that there is only scarce (if any) evidence for recoiling MBHs in {\it Sloan Digital Sky Survey} quasars.

\begin{figure}[b]
\centering
\includegraphics[width=.48\textwidth]{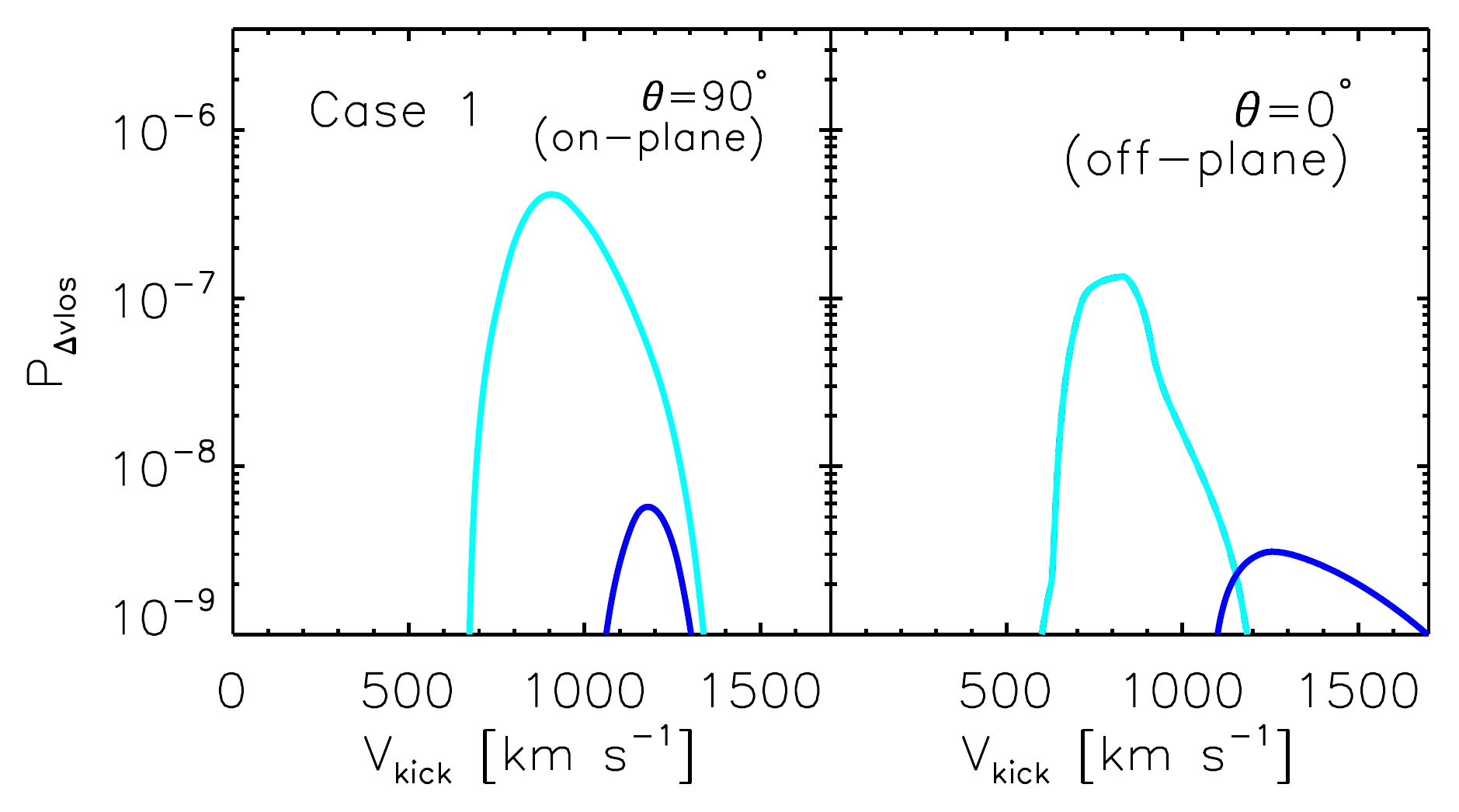}
\caption{Probability of observing a recoiling MBH in the major merger remnant as a kinematically-offset AGN. \textit{Cyan curves:} probability that the MBH
has a velocity offset along the line of sight $\Delta v_{\rm los}\ge 600\,\kms$ while shining with an optical luminosity $\ge L_{t,B}$ (for $z=0.1$). 
\textit{Blue curves:} same for $\Delta v_{\rm los} \ge 1,000\,\kms$. Both probabilities were computed assuming our best-case scenario Case 1 (see text for details).}

\label{prob_case1_dv}
\vskip 0.3cm
\end{figure}

To assess the detectability of recoiling MBHs as kinematically offset AGNs, we have generated a library of orbits using the semi-analytical model described in Section~\ref{model}, 
for kick velocities lower than the escape speed from the host and return timescales shorter than the Hubble time. We have computed the accretion rate and the resulting optical
and X-ray luminosities along the trajectory of the recoiling holes. 

Figure~\ref{timescales_dv} shows the return timescales of recoiling MBHs in the major merger remnant, for different kick velocities in the plane of the nuclear disk
and perpendicular to it. The figure also shows the lifetime of the initial disk carried by the hole, the total time spent by the hole above the threshold 
velocity offsets $\Delta v_t \ge 600\kms$ and $\Delta v_t \ge 1,000\,\kms$, as well as the detectability window as a velocity-offset AGN, i.e. the total time 
during which the MBH is both above the velocity and the luminosity thresholds (at $z=0.1$). 
The detectability window of a recoiling hole depends on the kinematic profile of its orbit. For moderate kick velocities, where the MBH can undergo 
many pericenter passages before the initial disk is exhausted, more eccentric orbits yield higher relative velocities, making the MBH observable as a kinematically-offset AGN. 
When the MBH is kicked in the direction perpendicular to the disk, the return timescale is longer and orbits are less eccentric, yielding generally shorter 
observability windows for large kinematic displacements. In our major merger, a MBH launched with recoil velocity $\vkick=1,200\,\kms$ in 
the $\theta=90^{\circ}$ direction, will return to the center of the galaxy in about 0.3 Gyr, and would be observable as a kinematically-offset AGN for only 
0.2 Myr. The detectability window decreases by a factor of 4 in the $\theta=0^{\circ}$ case for the same recoil speed. In the minor mergers the threshold 
velocity is reached only for kicks with $\vkick>v_{\rm esc}$, which are are not shown here.

The probability of observing a recoiling hole as a velocity-offset AGN with $\Delta v_{\rm los}$ along the line of sight (LOS) can be written as 
\begin{eqnarray}
P_{\Delta v_{\rm los}}(\vec{V}_{\rm kick}) &=& P_{\{\Delta v_{\rm los} > \Delta v_t, L> L_t\}}(\vec{V}_{\rm kick},z) \\
\nonumber &\times & P_{\vkick} (\vec{a}_1,\vec{a}_2,q),
\end{eqnarray}
where $ P_{\{\Delta v_{\rm los} > \Delta v_t, L> L_t\}}(\vec{V}_{\rm kick})$ is the probability that a hole with initial kick $\vkick$ has an velocity offset 
$\Delta v_{\rm los} > \Delta v_t$ while moving along the LOS at the time of observation and shining with bolometric luminosity above the threshold $L_t$ (assuming no obscuration). 
Here, $P_{\vkick} (\vec{a}_1,\vec{a}_2,q)$ is the probability that a coalescing MBH binary with spin vectors $\vec{a_1}, \vec{a_2}$ and mass ratio $q$ recoils with 
velocity $\vkick > \Delta v_t$. In terms of the relevant timescales, 
\begin{eqnarray}\label{Pdv}
P_{\Delta v_{\rm los}}(\vec{V}_{\rm kick},z) &=& \left [ \frac{ t_{\{ \Delta v_{\rm los} \ge v_t, L \ge L_t\}}(\vec{V}_{\rm kick}) } {t_{\rm return}} \right] \\
\nonumber & \times &  P_{\vkick} (\vec{a}_1,\vec{a}_2,q),
\end{eqnarray}
where $t_{\{ \Delta v_{\rm los} \ge v_t, L \ge L_t\}}(\vec{V}_{\rm kick},z)$ is the integrated time during which $\Delta v_{\rm los} > \Delta v_t$ along the LOS, 
and $L \ge L_t$. 

We have evaluated equation (\ref{Pdv}) for a range of recoil velocities, converting each value of $\vkick$ into an effective recoil velocity $V_{\rm eff}$ (using Equation ~\ref{veff}),
and interpolating the values of $P_{\{\Delta v_{\rm los} > \Delta v_t, L> L_t\}}$ using the timescales shown in Figure~\ref{timescales_dv}. The resulting probability 
is shown in Figure~\ref{prob_case1_dv} under the assumptions of our best-case scenario Case 1 (random alignment of the binary orbital plane and spin).
The probability peaks at $\sim 800\,\kms$ to a maximum value of only $P_{\Delta v_{\rm los}}\sim 5\times10^{-7}$ for $\Delta v_{\rm los} \ge 600\,\kms$, and is nearly 
2 orders of magnitude smaller for $\Delta v_{\rm los} \ge 1,000\,\kms$. We calculated the same probability in Case 2, where the spins of the binary are aligned 
with the orbital angular momentum vector, the orbital plane is aligned with the gaseous disk, and all kicks occur in the plane of the disk. 
In this scenario the maximum recoil velocity is $\vkick\sim200\,\kms$ and the holes are unobservable as kinematically-offset AGNs. 

\begin{figure*}[h]
\centering
\includegraphics[width=.7\textwidth]{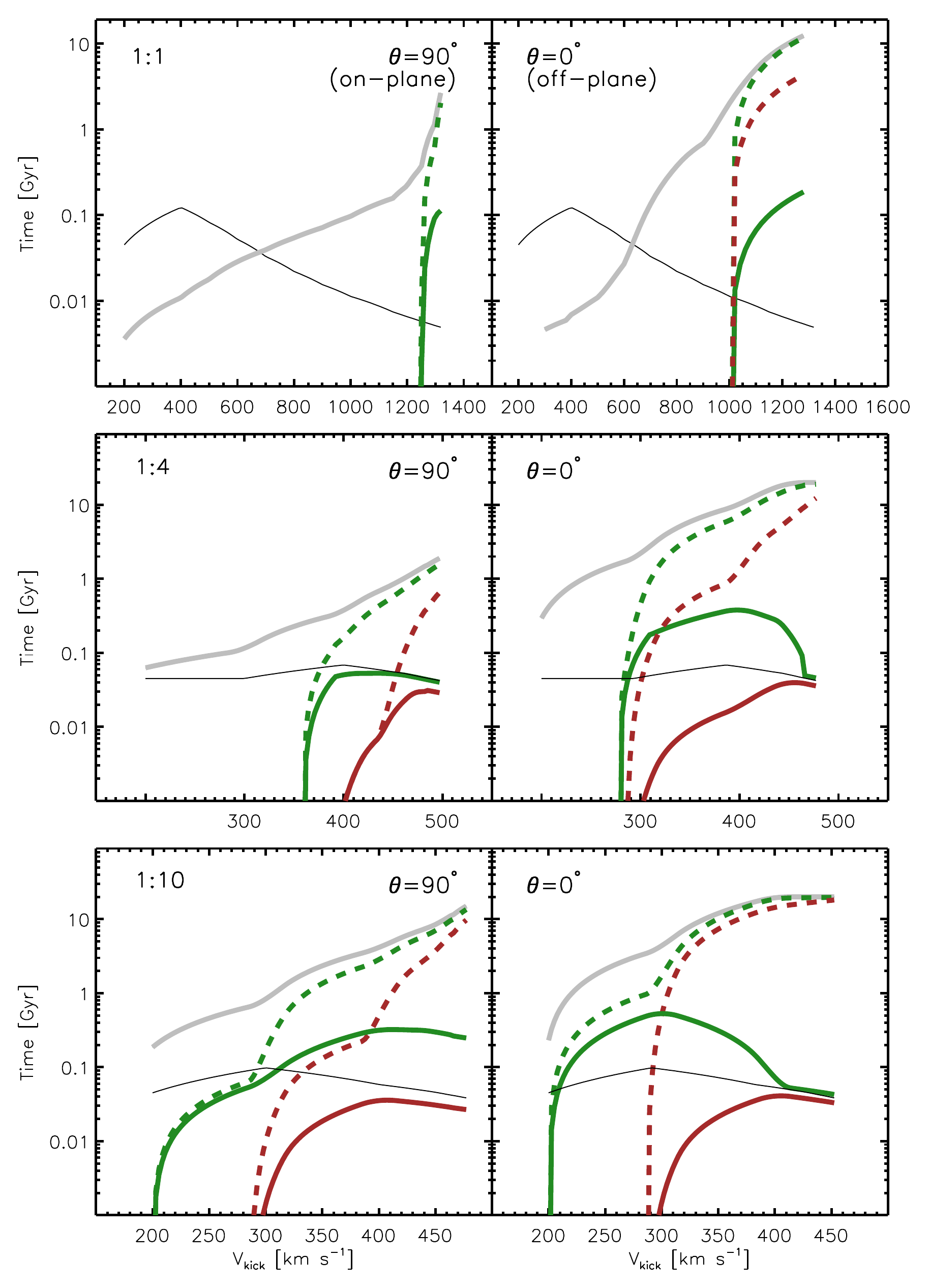}
\caption{Relevant timescales for assessing the detectability of recoiling MBHs as spatially-offset AGNs at $z=0.1$, as a function of the 
kick velocity. \textit{Gray curves}: return timescale of a MBH recoiling in the 1:1, 1:14, and 1:10 merger remnants.
The displacement threshold is set to $\Delta r_t = 0.4"$. \textit{Dashed green curves:} total time during which the MBH has a spatial offsets of $\Delta r \ge \Delta r_t=0.4"$.
\textit{Solid green curves:} total time during which the MBH has a spatial offsets of $\Delta r \ge \Delta r_t=0.4"$ while shinning with luminosity $L\ge L_{t,B}$. 
\textit{Dashed brown lines} and \textit{solid brown lines:} same total times but for the thresholds $\Delta r_t = 2"$ and $L\ge L_{t,X}$. 
\textit{Solid black line:}  the lifetime of the initial disk as a function of recoil velocity.}
\label{timescales_dx}
\vskip 0.3cm
\end{figure*}
%

\subsection{Spatially-offset AGNs}\label{dx}

We can use a similar method to that described above to estimate the observability of recoling MBHs as spatially-offest AGNs. 
Figure~\ref{timescales_dx} shows the integrated time during which the MBH is displaced by a threshold distance of $\Delta r_t=0.4"$ and $\Delta r_t=2"$ at $z=0.1$ 
(corresponding to twice the resolution of \textit{HST/JWST} and \textit{Chandra}, respectively), as well as the integrated time during which the MBH is both 
displaced by $\Delta r_t$ and is shining with $L\ge L_t$.  The upper, middle, and lower panels show results for the 1:1, 1:4, and 1:10 mergers, respectively.

In the major merger, only large recoil velocities (of the order the escape speed) can displace the MBH by $\Delta r_t$ when the kick occurs in the plane of the gaseous disk. 
The hole can shine for timescales far longer than the lifetime of the accretion disk carried along, since there is a large supply of gas in the galactic disk for fueling 
near apocenter, where the MBH spends most of its wandering time. 

In the minor mergers, the potential wells are shallower and large apocenter distances can be reached at low kick velocities ($\vkick < 500\,\kms$).

For trajectories having low inclinations relative to the gaseous disk, Bondi accretion can be significant and the MBH can wander sufficiently far from the center of 
the host to be detectable as an offset X-ray AGN at $z\lta 0.1$. The detectability window typically ranges from a few to tens of Myr. 

The probability of observing a recoiling MBH as a spatially-offset AGN with displacement $\Delta r_p$ perpendicular to the LOS can be written as 
\begin{eqnarray}
P_{\Delta r_p}(\vec{V}_{\rm kick},z) &=& P_{\{\Delta r_p > \Delta r_t, L> L_t\}}(\vec{V}_{\rm kick},z) \\
\nonumber &\times & P_{\vkick} (\vec{a}_1,\vec{a}_2,q),
\end{eqnarray}
where $P_{\{\Delta r_p > \Delta r_t, L> L_t\}}$ is the probability that the hole is shining with resulting luminosity $L>L_t$ and is displaced by a projected 
distance larger than the threshold $\Delta r_t$. In terms of the relevant timescales,
\begin{eqnarray}\label{px}
P_{\Delta r_p}(\vec{V}_{\rm kick},z) &=& \left [ \frac{ t_{\{ \Delta r_p \ge r_t, L \ge L_t\}}(\vec{V}_{\rm kick},z) } {t_{\rm return}} \right] \\
\nonumber & \times &  P_{\vkick} (\vec{a}_1,\vec{a}_2,q).
\end{eqnarray}
%

\begin{figure}[t]
\centering
\includegraphics[width=.48\textwidth]{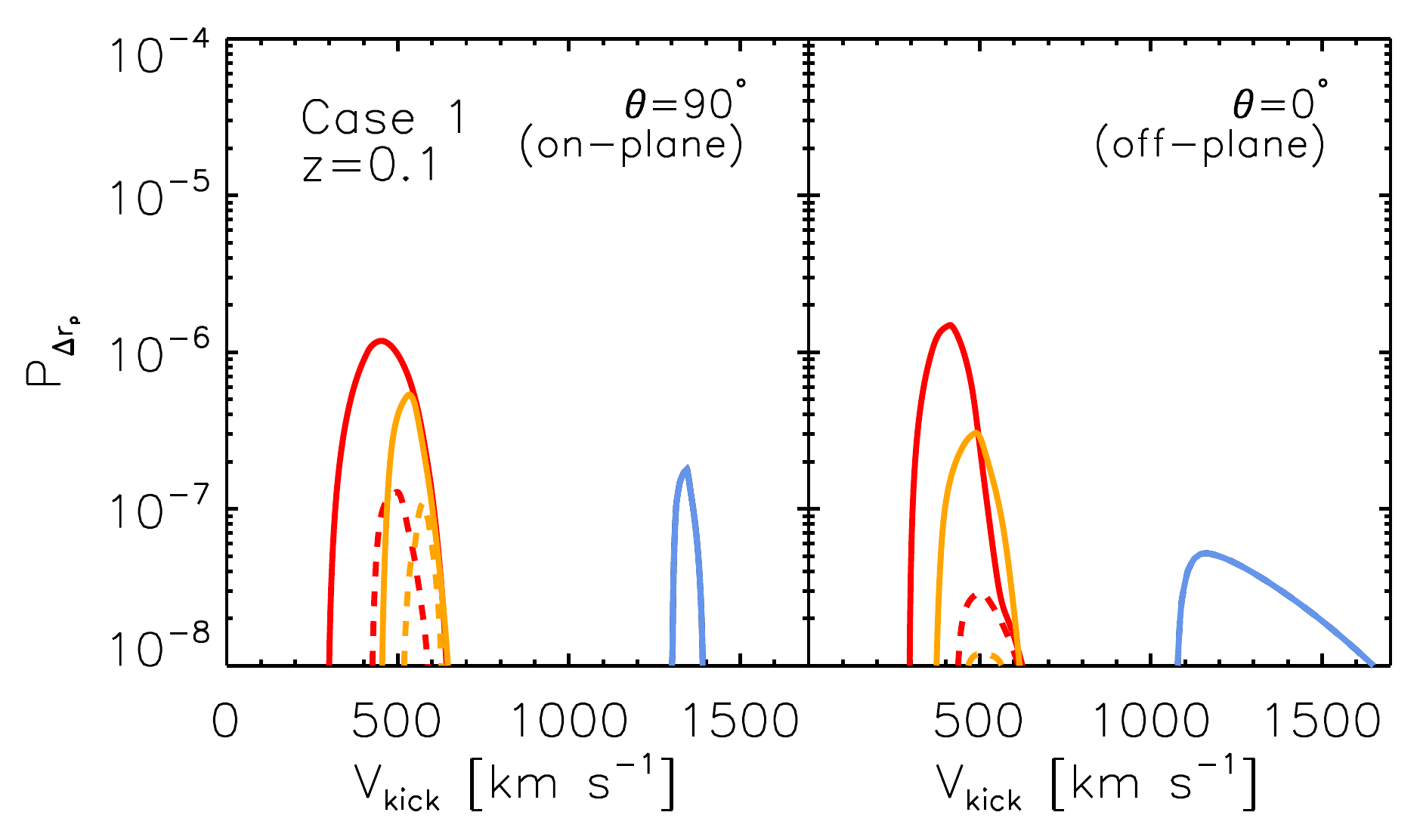}
\caption{Probability of observing a recoiling MBH as a spatially-offset AGN. The colors represent the 1:1 (\textit{blue}), 1:4 (\textit{orange}), and 1:10 (\textit{red}) merger 
remnants. \textit{Solid lines:} detectability thresholds of $\Delta r_t=0.4"$ and $L\ge L_{t,B}$. \textit{Dashed lines:} $\Delta r_t =2"$ and $L\ge L_{t,X}$.
All probabilities were computed assuming our best-case scenario Case 1 (see text for details).}
\label{prob_case1_dr}
\vskip 0.3cm
\end{figure}

Figure~\ref{prob_case1_dr} shows the probability $P_{\Delta r_p}$ that a MBH could be observed as a spatially-offset AGN during its wandering time assuming Case 1. 
Along both orientations considered here (on and off the plane of the gaseous disk), the probability peaks at $\sim 10^{-6}$ for optical detection, $\vkick \sim 500\,\kms$, and is higher 
for the 1:10 merger where the recoiling hole reaches larger apocenter distances. The probability that such a source could be observed by \textit{Chandra} instead 
drops by nearly an order of magnitude. 
For the 1:1 merger remnant the peak of the distribution is $\sim 10^{-6}$ in the $\theta=90^{\circ}$ case.  The distribution broadens in the $\theta=0^{\circ}$ case 
peaks at a lower value since the MBH spends a considerable fraction of its return time in a non-active phase. 
We have carried out the same calculation in Case 2, and find that the likelyhood of observing spatially-offset AGNs becomes negligible. There are several reasons for 
this: (1) when spins are aligned, the maximum recoil velocity is $\sim 200\,\kms$; (2) at low recoil speeds the mass carried by the disk is $M_{\rm disk}=M_\bullet$, and 
the effective recoil velocity is $\vkick/2$; and (3) recoil velocities in the direction perpendicular to the gaseous disk are not allowed since $\alpha_{1,2}=0$.  

\section{Summary and Discussion}\label{summary}

We have studied the orbital evolution of recoiling MBHs in post-merger environments via high-resolution $N$-body + SPH simulations and a semi-analytical model 
that combines the results of merger simulations with a prescription for MBH accretion. Our results can be summarized as follows:

\begin{enumerate}

\item We find little evidence for an observable imprint of a recoiling, non-active MBH in the gaseous turbulent medium of our major merger remnant. In our simulation, 
the total energy deposited by the moving hole is only $2\%$ of the change of internal energy of the gas, even at high Mach numbers. 
The contribution to the thermodynamics of the ambient gas becomes significant in minor merger remnants instead, potentially allowing for an electromagnetic signature of MBH recoil.

\item We have built a library of MBH orbits in three gas rich galaxy mergers, allowing the holes to carry a punctured accretion disk and the associated BLR upon ejection,
and to accrete from the ISM when the disk is exhausted. In the major merger remnant, the combination of both deeper central potential well and drag from 
high-density gas confines even MBHs with kick velocities as high as $1,200\,\kms$ to within 1 kpc from the host's center. In minor merger remnants gas drag
is weaker, the potential is shallower, and smaller recoil speeds can displace the hole to larger distances. 

\item We have dissected the orbits of recoiling holes into regions where they may be detectable as kinematically and/or or spatially offset AGNs. Kinematically-offset AGNs carry their 
accretion disk and associated BLR for a timescale $M_{\rm disk}/\dot M$ and can be observable as such when their broad emission lines have a velocity shift $> 600\,\kms$ relative 
to the narrow-line gas left behind. In the major merger remnant, moderate recoil velocities ($700-1,000\,\kms$) provide the most likely scenarios for the detection of kinematically 
offset nuclei, as the initial disk can fuel the MBH during several pericenter (high-velocity) passages. The convolution of the total fractional time spent as a kinematically-offset 
AGN with the probability that a coalescing binary receives a given kick velocity yields the overall probability of observing recoiling MBHs as kinematically-offset AGNs.   
This is small, of order $\sim 10^{-7}$, even in the optimistic case (our ``Case 1") where the spins of the binary holes are randomly oriented relative to the orbital angular momentum 
vector and the orbital plane of the binary is randomly oriented relative to the galactic disk. It is totally negligible when the spins are aligned instead 
\citep[e.g.][]{bogdanovic07}, and the maximum recoil speed drops to $\sim 200\,\kms$. 

\item At low redshift, spatially-offset AGN can be observed with both \textit{Chandra} and \textit{HST/JWST} for a few Myr in major mergers for high recoil speeds 
($>1,000\,\kms$). In minor mergers, the detectability window shifts to lower recoil speeds ($\vkick < 500\,\kms$) and extends to up to a few 100 Myr in some cases. 
Under the Case 1 optimistic scenario, the probability of observing spatially-offset AGNs is larger (by two dex) in the 1:10 merger than in the 1:1 merger, since in the 
former larger offsets occur for lower, more likely kicks. The peak of the probability distribution ($10^{-6}$) occurs for $\vkick \simeq 500\,\kms$ and optical detection,
and is lower by about one dex in the case of X-ray detection. 

\end{enumerate}

Our semi-analytical model has been calibrated using a limited set of numerical simulations. The choice of a gas fraction of $10\%$ for the major merger is rather conservative; 
a larger gas content, quite likely at high redshift, would cause increased drag on the MBH and reinforce our conclusions. For the 1:10 minor merger, representative
of some of the most common types of mergers during the hierarchical assembly of galaxies, we have assumed a larger gas fraction, 30\%, as in this case the pairing of the 
two holes and their coalescence is significantly faster. Such a choice is still on the conservative side; the typical gas fraction in galaxies with masses in 
the range $2 \times 10^{10} - 2 \times 10^{11} M_{\odot}$ (corresponding to the virial masses of the primary and secondary galaxies in the 1:10 mergers, respectively)
is in the range 30-50\% at low redshift \citep{simon07, leroy08} and would be even higher at early epochs. As for the mass ratio, our simulations sample quite well 
the regime in which the dynamical friction timescale is short enough for the galaxies to merge in a few orbits. 
At larger mass ratios the satellite galaxies will suffer severe tidal mass losses as dynamical friction reduces their pericentre distance: this will increase 
substantially their decay time \citep{colpi99,taffoni03} and inhibit the formation of a MBH binary. 
  
A limitation of the present study is that our simulations do not capture all the phases of the ISM: gas is 
either warm, with temperatures close to $10^4$ K, or hot ($T \sim 10^5 - 10^6$ K) as a result of heating by supernovae
explosions or shocks in the final collision of the merging galaxy cores. The cold molecular phase that is believed to represent a significant 
fraction of the mass of the nuclear disk \citep[e.g.][]{Downes98} is missing in our current calculations. At temperatures of 
hundreds of kelvins in the presence of a starburst \citep{klessen07}, molecular material may be arranged in a thin and dense sheet 
the midplane of the nuclear disk \citep[e.g.][]{wada01}. MBHs recoiling along the plane may then suffer more dynamical friction, and
one may expect a greater difference between the fate of in-plane and off-plane recoils than seen in our study.
 
The finite resolution of our numerical simulations may also affect our results. In particular, the resolution of our 1:4 merger (200 pc) results in a sharp decline 
of the gas distribution in the remnant nucleus. To quantify this effect, we have rescaled the central gas and stellar densities to their corresponding values at 
200 pc. We find that the recoiling MBHs in the 1:4 merger reach smaller apocenter distances and wander for even shorter timescales. The effect is
more dramatic for kicks below $200\,\kms$. Since MBHs with these velocities  are not observable as neither kinematically nor spatially-offset AGNs, 
our detection probabilities remain largely unaffected.

Another important limitation of our semi-analytical modelling is the neglect of AGN feedback, which acts to reduce the efficiency of frictional drag by
heating  the gas surrounding the recoiling MBH. 
Recent work by \cite{sijacki2010} shows that in simulations where AGN feedback is
fully included, the MBH creates a hot low-density (instead of a high-density) wake that expands away  from the hole. The MBH describes a precessing
elliptical orbit that tend to circularize, and decays towards the center of the remnant on a longer return timescale.
To get a simple estimate of the effect of AGN feedback on our calculations we have assumed in our semi-analytic
model that at each time-step a fraction equal to 5\%  of the radiative energy produced by the accreting MBH  goes directly towards increasing the local thermal energy of the gas.
We find that in dense regions where the hole  is accreting  at high Bondi rates, the reduction of the effective Mach number from such energy  deposition lengthen the return timescale by 15\% in the major merger for $\vkick=1000\kms$ and 20\% for $\vkick=500\kms$ in the 1:10 merger. The time during which the MBHs are observable as spatially offset AGN increases by 6\% and 9\% respectively.  The overall detection probability however, remains low.

Furthermore, in our detectability calculations we have not included obscuration by dust. Because theoptical and X-ray reddening does not depend on the orientation of the 
host galaxy \citep{winter10}, it is believed that most of the obscuration occurs in a region close to the MBH, with optical light being more strongly affected that X-rays 
and infrared wavelengths \citep[e.g.][]{goulding09,winter10}. In particular, the recent study of \cite{koss10} shows that the \textit{Swift} BAT X-ray survey has 
revealed roughly 5-6 times more dual AGNs than the Sloan Digital Sky Survey (SDSS) due to optical extinction and dilution by star formation. As a result, the detection 
probability is further reduced for \textit{HST/JWST} sources. 

While we have pointed out the low likelihood that moderate-to-large recoils can be observed, one possibly observable consequence of this phenomenon is the existence of a population of hot-dust-poor AGNs due to the removal of the MBH and associated BLR from the surrounding hot dust region. Nearly $10\%$ of  spectroscopically confirmed type 1 AGNs samples \citep{elvis94,richards06,hao10a}, and XMM-COSMOS (Elvis et al. 2010 in prep) show a relatively weak infrared bump, associated with dust emission. The number of these objects has been shown to increase with redshift, from 6\% at $z<2$ to $20\%$ at $2<z<3.5$ \citep{hao10b}. They find that the covering factor of these AGN is only 2-30\%, significantly smaller than the 75\% predicted by the unified model. Since these AGN are in the redshift range $0\le z \le4$, these hot-dust-free AGN are not recently born AGN which have not time to form a dusty torus \citep{jiang10}. Based on their spectral resolution (2 \AA), a $z=2$ source could have a relative velocity of $50-70\kms$.

\acknowledgements
We thank the anonymous referee for thoughtful comments that substantially improved our paper. This research was funded by NASA through grants NNX09AJ34G (P.M.) and by NSF 
through a graduate student research fellowship (J.G). Simulations were carried  out on the Pleiades supercomputer at NASA Ames, and some results were rendered using the 
SPLASH code \citep{price07}. 

\bibliographystyle{apj}

\begin{thebibliography}{83}
\expandafter\ifx\csname natexlab\endcsname\relax\def\natexlab#1{#1}\fi

\bibitem[{Baker {et~al.}(2008)Baker, Boggs, Centrella, Kelly, McWilliams,
  Miller, \& van Meter}]{baker08}
Baker, J.~G., Boggs, W.~D., Centrella, J., Kelly, B.~J., McWilliams, S.~T.,
  Miller, M.~C., \& van Meter, J.~R. 2008, \apj, 682, L29

\bibitem[{Baker {et~al.}(2006)Baker, Centrella, Choi, Koppitz, van Meter, \&
  Miller}]{Baker06}
Baker, J.~G., Centrella, J., Choi, D.-I., Koppitz, M., van Meter, J.~R., \&
  Miller, M.~C. 2006, \apj, 653, L93

\bibitem[{Begelman {et~al.}(1980)Begelman, Blandford, \& Rees}]{begelman80}
Begelman, M.~C., Blandford, R.~D., \& Rees, M.~J. 1980, Nature, 287, 307

\bibitem[{Bekenstein(1973)}]{Bekenstein73}
Bekenstein, J.~D. 1973, \apj, 183, 657

\bibitem[{Blecha \& Loeb(2008)}]{blecha08}
Blecha, L., \& Loeb, A. 2008, \mnras, 390, 1311

\bibitem[{Bogdanovi{\'c} {et~al.}(2007)Bogdanovi{\'c}, Reynolds, \&
  Miller}]{bogdanovic07}
Bogdanovi{\'c}, T., Reynolds, C.~S., \& Miller, M.~C. 2007, \apj, 661, L147

\bibitem[{Bonning {et~al.}(2007)Bonning, Shields, \& Salviander}]{bonning07}
Bonning, E.~W., Shields, G.~A., \& Salviander, S. 2007, \apj, 666, L13

\bibitem[{Callegari {et~al.}(2010)Callegari, Kazantzidis, Mayer, Colpi,
  Bellovary, Quinn, \& Wadsley}]{callegari10}
Callegari, S., Kazantzidis, S., Mayer, L., Colpi, M., Bellovary, J.~M., Quinn,
  T., \& Wadsley, J. 2010, eprint arXiv, 1002, 1712

\bibitem[{Callegari {et~al.}(2009)Callegari, Mayer, Kazantzidis, Colpi,
  Governato, Quinn, \& Wadsley}]{callegari09}
Callegari, S., Mayer, L., Kazantzidis, S., Colpi, M., Governato, F., Quinn, T.,
  \& Wadsley, J. 2009, \apjl, 696, L89

\bibitem[{Campanelli {et~al.}(2007{\natexlab{a}})Campanelli, Lousto, Zlochower,
  \& Merritt}]{Campanelli2007}
Campanelli, M., Lousto, C., Zlochower, Y., \& Merritt, D. 2007{\natexlab{a}},
  \apj, 659, L5

\bibitem[{Campanelli {et~al.}(2006)Campanelli, Lousto, Marronetti, \&
  Zlochower}]{Campanelli06}
Campanelli, M., Lousto, C.~O., Marronetti, P., \& Zlochower, Y. 2006, \prl, 96,
  111101

\bibitem[{Campanelli {et~al.}(2007{\natexlab{b}})Campanelli, Lousto, Zlochower,
  \& Merritt}]{Campanelli07}
Campanelli, M., Lousto, C.~O., Zlochower, Y., \& Merritt, D.
  2007{\natexlab{b}}, \prl, 98, 231102

\bibitem[{Chandrasekhar(1943)}]{chandrasekhar43}
Chandrasekhar, S. 1943, \apj, 97, 255

\bibitem[{Civano {et~al.}(2010)Civano, Elvis, Lanzuisi, Jahnke, Zamorani,
  Blecha, Bongiorno, Brusa, Comastri, Hao, Leauthaud, Loeb, Mainieri,
  Piconcelli, Salvato, Scoville, Trump, Vignali, Aldcroft, Bolzonella,
  Bressert, Finoguenov, Fruscione, Koekemoer, Cappelluti, Fiore, Giodini,
  Gilli, Impey, Lilly, Lusso, Puccetti, Silverman, Aussel, Capak, Frayer,
  Floch, McCracken, Sanders, Schiminovich, \& Taniguchi}]{civano10}
Civano, F., {et~al.} 2010, \apj, 717, 209

\bibitem[{Collin \& Zahn(1999)}]{collin99}
Collin, S., \& Zahn, J.-P. 1999, Astrophysics and Space Science, 265, 501

\bibitem[{Colpi {et~al.}(1999)Colpi, Mayer, \& Governato}]{colpi99}
Colpi, M., Mayer, L., \& Governato, F. 1999, \apj, 525, 720

\bibitem[{Comerford {et~al.}(2009)Comerford, Gerke, Newman, Davis, Yan, Cooper,
  Faber, Koo, Coil, Rosario, \& Dutton}]{comerford09a}
Comerford, J.~M., {et~al.} 2009, \apj, 698, 956

\bibitem[{Davies {et~al.}(2004)Davies, Tacconi, \& Genzel}]{Davies04}
Davies, R.~I., Tacconi, L.~J., \& Genzel, R. 2004, \apj, 613, 781

\bibitem[{Devecchi {et~al.}(2009)Devecchi, Rasia, Dotti, Volonteri, \&
  Colpi}]{devecchi09}
Devecchi, B., Rasia, E., Dotti, M., Volonteri, M., \& Colpi, M. 2009, \mnras,
  394, 633

\bibitem[{Dotti {et~al.}(2007)Dotti, Colpi, Haardt, \& Mayer}]{dotti07}
Dotti, M., Colpi, M., Haardt, F., \& Mayer, L. 2007, \mnras, 379, 956

\bibitem[{Downes \& Solomon(1998)}]{Downes98}
Downes, D., \& Solomon, P.~M. 1998, \apj, 507, 615

\bibitem[{Elvis {et~al.}(1994)Elvis, Wilkes, McDowell, Green, Bechtold,
  Willner, Oey, Polomski, \& Cutri}]{elvis94}
Elvis, M., {et~al.} 1994, The Astrophysical Journal Supplement Series, 95, 1

\bibitem[{Escala {et~al.}(2004)Escala, Larson, Coppi, \& Mardones}]{escala04}
Escala, A., Larson, R.~B., Coppi, P.~S., \& Mardones, D. 2004, \apj, 607, 765

\bibitem[{Escala {et~al.}(2005)Escala, Larson, Coppi, \& Mardones}]{escala05}
---. 2005, \apj, 630, 152

\bibitem[{Fakhouri \& Ma(2008)}]{Fakhouri08}
Fakhouri, O., \& Ma, C.-P. 2008, \mnras, 386, 577

\bibitem[{Fitchett \& Detweiler(1984)}]{Fitchett84}
Fitchett, M.~J., \& Detweiler, S. 1984, Royal Astronomical Society, 211, 933

\bibitem[{Gerke {et~al.}(2007)Gerke, Newman, Lotz, Yan, Barmby, Coil,
  Conselice, Ivison, Lin, Koo, Nandra, Salim, Small, Weiner, Cooper, Davis,
  Faber, \& Guhathakurta}]{gerke07}
Gerke, B.~F., {et~al.} 2007, \apj, 660, L23

\bibitem[{Gonz{\'a}lez {et~al.}(2007)Gonz{\'a}lez, Hannam, Sperhake,
  Br{\"u}gmann, \& Husa}]{Gonzalez2007}
Gonz{\'a}lez, J.~A., Hannam, M., Sperhake, U., Br{\"u}gmann, B., \& Husa, S.
  2007, \prl, 98, 231101

\bibitem[{Goodman \& Tan(2004)}]{goodman04}
Goodman, J., \& Tan, J.~C. 2004, \apj, 608, 108

\bibitem[{Goulding \& Alexander(2009)}]{goulding09}
Goulding, A.~D., \& Alexander, D.~M. 2009, \mnras, 398, 1165

\bibitem[{Gualandris \& Merritt(2008)}]{gualandris08}
Gualandris, A., \& Merritt, D. 2008, \apj, 678, 780

\bibitem[{Guedes {et~al.}(2008)Guedes, Diemand, Zemp, Kuhlen, Madau, \&
  Mayer}]{guedes08b}
Guedes, J., Diemand, J., Zemp, M., Kuhlen, M., Madau, P., \& Mayer, L. 2008,
  Astron. Naschr., 329, 1004

\bibitem[{Guedes {et~al.}(2009)Guedes, Madau, Kuhlen, Diemand, \&
  Zemp}]{guedes09}
Guedes, J., Madau, P., Kuhlen, M., Diemand, J., \& Zemp, M. 2009, \apj, 702,
  890

\bibitem[{Haehnelt(1994)}]{haehnelt94}
Haehnelt, M.~G. 1994, \mnras, 269, 199

\bibitem[{Hao {et~al.}(2010{\natexlab{a}})Hao, Elvis, Civano, Lanzuisi, Brusa,
  Lusso, Zamorani, Comastri, Bongiorno, Impey, Koekemoer, Floc'h, Salvato,
  Sanders, Trump, \& Vignali}]{hao10a}
Hao, H., {et~al.} 2010{\natexlab{a}}, \apjl, 724, L59

\bibitem[{Hao {et~al.}(2010{\natexlab{b}})Hao, Elvis, Civano, \&
  Lawrence}]{hao10b}
Hao, H., Elvis, M., Civano, F., \& Lawrence, A. 2010{\natexlab{b}}, eprint
  arXiv, 1011, 429, 5 pages, 4 figures, ApJL submitted

\bibitem[{Hernquist(1990)}]{Hernquist90}
Hernquist, L. 1990, \apj, 356, 359

\bibitem[{Herrmann {et~al.}(2007)Herrmann, Hinder, Shoemaker, Laguna, \&
  Matzner}]{Herrmann07}
Herrmann, F., Hinder, I., Shoemaker, D., Laguna, P., \& Matzner, R.~A. 2007,
  \apj, 661, 430

\bibitem[{Jiang {et~al.}(2010)Jiang, Fan, Brandt, Carilli, Egami, Hines, Kurk,
  Richards, Shen, Strauss, Vestergaard, \& Walter}]{jiang10}
Jiang, L., {et~al.} 2010, Nature, 464, 380

\bibitem[{King \& Pringle(2006)}]{king06}
King, A.~R., \& Pringle, J.~E. 2006, \mnras, 373, L90

\bibitem[{Klessen {et~al.}(2007)Klessen, Spaans, \& Jappsen}]{klessen07}
Klessen, R.~S., Spaans, M., \& Jappsen, A.-K. 2007, \mnras, 374, L29

\bibitem[{Komossa {et~al.}(2003)Komossa, Burwitz, Hasinger, Predehl, Kaastra,
  \& Ikebe}]{komossa03}
Komossa, S., Burwitz, V., Hasinger, G., Predehl, P., Kaastra, J.~S., \& Ikebe,
  Y. 2003, \apj, 582, L15

\bibitem[{Komossa {et~al.}(2008)Komossa, Zhou, \& Lu}]{Komossa08}
Komossa, S., Zhou, H., \& Lu, H. 2008, \apj, 678, L81

\bibitem[{Koss {et~al.}(2010)Koss, Mushotzky, Veilleux, \& Winter}]{koss10}
Koss, M., Mushotzky, R., Veilleux, S., \& Winter, L. 2010, \apjl, 716, L125

\bibitem[{Lacey \& Cole(1993)}]{Lacey93}
Lacey, C., \& Cole, S. 1993, Royal Astronomical Society, 262, 627

\bibitem[{Leroy {et~al.}(2008)Leroy, Walter, Brinks, Bigiel, de~Blok, Madore,
  \& Thornley}]{leroy08}
Leroy, A.~K., Walter, F., Brinks, E., Bigiel, F., de~Blok, W. J.~G., Madore,
  B., \& Thornley, M.~D. 2008, \aj, 136, 2782

\bibitem[{Liu {et~al.}(2010)Liu, Greene, Shen, \& Strauss}]{liu10}
Liu, X., Greene, J.~E., Shen, Y., \& Strauss, M.~A. 2010, \apjl, 715, L30

\bibitem[{Loeb(2007)}]{loeb07}
Loeb, A. 2007, \prl, 99, 41103

\bibitem[{Lousto {et~al.}(2010)Lousto, Nakano, Zlochower, \&
  Campanelli}]{lousto10}
Lousto, C.~O., Nakano, H., Zlochower, Y., \& Campanelli, M. 2010, \prd, 81,
  84023

\bibitem[{Madau \& Quataert(2004)}]{madau04}
Madau, P., \& Quataert, E. 2004, \apj, 606, L17

\bibitem[{Magain {et~al.}(2005)Magain, Letawe, Courbin, Jablonka, Jahnke,
  Meylan, \& Wisotzki}]{Magain05}
Magain, P., Letawe, G., Courbin, F., Jablonka, P., Jahnke, K., Meylan, G., \&
  Wisotzki, L. 2005, Nature, 437, 381

\bibitem[{Marconi {et~al.}(2004)Marconi, Risaliti, Gilli, Hunt, Maiolino, \&
  Salvati}]{marconi04}
Marconi, A., Risaliti, G., Gilli, R., Hunt, L.~K., Maiolino, R., \& Salvati, M.
  2004, \mnras, 351, 169

\bibitem[{Max {et~al.}(2007)Max, Canalizo, \& de~Vries}]{max07}
Max, C.~E., Canalizo, G., \& de~Vries, W.~H. 2007, Science, 316, 1877

\bibitem[{Mayer {et~al.}(2007)Mayer, Kazantzidis, Madau, Colpi, Quinn, \&
  Wadsley}]{mayer07}
Mayer, L., Kazantzidis, S., Madau, P., Colpi, M., Quinn, T., \& Wadsley, J.
  2007, Science, 316, 1874

\bibitem[{Merritt \& Milosavljevi{\'c}(2005)}]{merritt05}
Merritt, D., \& Milosavljevi{\'c}, M. 2005, Living Reviews in Relativity, 8, 8

\bibitem[{Mo {et~al.}(1998)Mo, Mao, \& White}]{mo98}
Mo, H.~J., Mao, S., \& White, S. D.~M. 1998, \mnras, 295, 319

\bibitem[{Ostriker(1999)}]{ostriker99}
Ostriker, E.~C. 1999, \apj, 513, 252

\bibitem[{Peres(1962)}]{Peres62}
Peres, A. 1962, Phys. Rev., 128, 2471

\bibitem[{Pierce {et~al.}(2010)Pierce, Lotz, Salim, Laird, Coil, Bundy,
  Willmer, Rosario, Primack, \& Faber}]{pierce10}
Pierce, C.~M., {et~al.} 2010, \mnras, 1285

\bibitem[{Pretorius(2005)}]{Pretorius05}
Pretorius, F. 2005, \prl, 95, 121101

\bibitem[{Price(2007)}]{price07}
Price, D.~J. 2007, Publications of the Astronomical Society of Australia, 24,
  159

\bibitem[{Richards {et~al.}(2006)Richards, Lacy, Storrie-Lombardi, Hall,
  Gallagher, Hines, Fan, Papovich, Berk, Trammell, Schneider, Vestergaard,
  York, Jester, Anderson, Budav{\'a}ri, \& Szalay}]{richards06}
Richards, G.~T., {et~al.} 2006, The Astrophysical Journal Supplement Series,
  166, 470

\bibitem[{Rossi {et~al.}(2010)Rossi, Lodato, Armitage, Pringle, \&
  King}]{rossi10}
Rossi, E.~M., Lodato, G., Armitage, P.~J., Pringle, J.~E., \& King, A.~R. 2010,
  \mnras, 401, 2021

\bibitem[{Sesana {et~al.}(2004)Sesana, Haardt, Madau, \& Volonteri}]{Sesana04}
Sesana, A., Haardt, F., Madau, P., \& Volonteri, M. 2004, \apj, 611, 623

\bibitem[{Sesana {et~al.}(2005)Sesana, Haardt, Madau, \& Volonteri}]{Sesana05}
---. 2005, \apj, 623, 23

\bibitem[{Shields {et~al.}(2009)Shields, Bonning, \& Salviander}]{shields09}
Shields, G.~A., Bonning, E.~W., \& Salviander, S. 2009, \apj, 696, 1367

\bibitem[{Sijacki {et~al.}(2010)Sijacki, Springel, \& Haehnelt}]{sijacki2010}
Sijacki, D., Springel, V., \& Haehnelt, M. 2010, eprint arXiv, 1008, 3313

\bibitem[{Simon \& Geha(2007)}]{simon07}
Simon, J.~D., \& Geha, M. 2007, \apj, 670, 313

\bibitem[{Smith {et~al.}(2010)Smith, Shields, Bonning, McMullen, Rosario, \&
  Salviander}]{smith10}
Smith, K.~L., Shields, G.~A., Bonning, E.~W., McMullen, C.~C., Rosario, D.~J.,
  \& Salviander, S. 2010, \apj, 716, 866

\bibitem[{Spaans \& Silk(2000)}]{spaans00}
Spaans, M., \& Silk, J. 2000, \apj, 538, 115

\bibitem[{Taffoni {et~al.}(2003)Taffoni, Mayer, Colpi, \&
  Governato}]{taffoni03}
Taffoni, G., Mayer, L., Colpi, M., \& Governato, F. 2003, \mnras, 341, 434

\bibitem[{Tanaka \& Haiman(2009)}]{tanaka09}
Tanaka, T., \& Haiman, Z. 2009, \apj, 696, 1798

\bibitem[{van Meter {et~al.}(2010)van Meter, Miller, Baker, Boggs, \&
  Kelly}]{vanMeter10}
van Meter, J.~R., Miller, M.~C., Baker, J.~G., Boggs, W.~D., \& Kelly, B.~J.
  2010, The Astrophysical Journal, 719, 1427

\bibitem[{Vasudevan \& Fabian(2007)}]{vasudevan2007}
Vasudevan, R.~V., \& Fabian, A.~C. 2007, \mnras, 381, 1235

\bibitem[{Vicari {et~al.}(2007)Vicari, Capuzzo-Dolcetta, \& Merritt}]{vicari07}
Vicari, A., Capuzzo-Dolcetta, R., \& Merritt, D. 2007, \apj, 662, 797

\bibitem[{Volonteri {et~al.}(2003)Volonteri, Haardt, \& Madau}]{volonteri03}
Volonteri, M., Haardt, F., \& Madau, P. 2003, \apj, 582, 559

\bibitem[{Volonteri \& Madau(2008)}]{volonteri08}
Volonteri, M., \& Madau, P. 2008, \apj, 687, L57

\bibitem[{Wada \& Norman(2001)}]{wada01}
Wada, K., \& Norman, C.~A. 2001, \apj, 547, 172

\bibitem[{Wadsley {et~al.}(2004)Wadsley, Stadel, \& Quinn}]{wadsley04}
Wadsley, J.~W., Stadel, J., \& Quinn, T. 2004, New Astronomy, 9, 137

\bibitem[{Winter {et~al.}(2010)Winter, Mushotzky, Lewis, Veilleux, Koss, \&
  Keeney}]{winter10}
Winter, L.~M., Mushotzky, R., Lewis, K., Veilleux, S., Koss, M., \& Keeney, B.
  2010, X-RAY ASTRONOMY 2009; PRESENT STATUS, 1248, 369

\bibitem[{Wyithe \& Loeb(2003)}]{wyithe03}
Wyithe, J. S.~B., \& Loeb, A. 2003, \apj, 590, 691

\bibitem[{Zakamska {et~al.}(2003)Zakamska, Strauss, Krolik, Collinge, Hall,
  Hao, Heckman, Ivezi{\'c}, Richards, Schlegel, Schneider, Strateva, Berk,
  Anderson, \& Brinkmann}]{zakamska03}
Zakamska, N.~L., {et~al.} 2003, \aj, 126, 2125

\bibitem[{Zhou {et~al.}(2004)Zhou, Wang, Zhang, Dong, \& Li}]{zhou04}
Zhou, H., Wang, T., Zhang, X., Dong, X., \& Li, C. 2004, \apj, 604, L33

\end{thebibliography}

\end{document}